\documentclass[aps,pre,twocolumn,groupedaddress,floatfix,longbibliography,superscriptaddress]{revtex4-1}

\usepackage{amsmath,amssymb,amsfonts,mathrsfs,bm,bbold,braket,slashed,cancel}

\usepackage{graphicx}
\usepackage{epsfig}
\usepackage{float}
\usepackage{subfigure}
\usepackage{multirow}
\usepackage{feynmf}

\usepackage{times}

\usepackage[usenames,dvipsnames,svgnames]{xcolor}

\usepackage{hyperref}
\hypersetup{
    colorlinks=true,
    linkcolor=Blue,
    citecolor=Blue,
    urlcolor=Blue
}

\usepackage{accents}
\usepackage{ulem}
\usepackage{simpler-wick}
\usepackage{orcidlink}

\DeclareUnicodeCharacter{039B}{\ensuremath{\Lambda}}

\newcommand{\bea}{\begin{eqnarray}}
\newcommand{\eea}{\end{eqnarray}}
\newcommand{\be}{\begin{equation}}
\newcommand{\ee}{\end{equation}}

\renewcommand\vec{\bm}

\newcommand{\nn}{\nonumber}
\newcommand{\ii}{\mathrm{i}}

\newcommand{\gn}{\gamma^\nu}
\newcommand{\gm}{\gamma^\mu}

\newcommand{\Op}[1]{\mathcal{O}^{(#1)}}

\newcommand{\C}{\mathcal{C}}

\newcommand{\wph}{\omega}

\begin{document}

\title{Dilepton Production in a Rotating Thermal Medium: The Rigid Rotation Approximation}
\author{Jorge David Casta\~no-Yepes~\orcidlink{0000-0002-8654-1304}}
\email{jorge.yepes@correounivalle.edu.co}
\affiliation{ Departamento de Física, Universidad del Valle, Ciudad Universitaria Meléndez,
Santiago de Cali 760032, Colombia}
\author{Enrique Mu\~noz~\orcidlink{0000-0003-4457-0817}}
\email{ejmunozt@uc.cl}
\affiliation{Facultad de F\'isica, Pontificia Universidad Cat\'olica de Chile, Vicu\~{n}a Mackenna 4860, Santiago, Chile}
\affiliation{Center for Nanotechnology and Advanced Materials CIEN-UC, Avenida Vicuña Mackenna 4860, Santiago, Chile}

\begin{abstract}
We investigate dilepton production in a thermalized quark--gluon plasma subject to global rotation, in the rigid rotating approximation. We consider a generic process involving quark-antiquark annihilation followed by the emission of a highly energetic virtual photon decaying into a dilepton pair. For this process, we compute the dilepton emission rate from the imaginary part of the photon polarization tensor, at finite temperature and vorticity. Our results show that vorticity induces characteristic modifications in the light dilepton channel, namely $e^-e^+$ production, where the emission spectrum exhibits a suppression at low transverse mass together with a mild shift of the production threshold. This behavior originates from the role of vorticity as an effective spin-dependent chemical potential that alters the available phase-space distribution for the emission process. In contrast, the $\mu^-\mu^+$ channel is {more weakly affected by} the rotational background, thus remaining dominated by its intrinsic mass threshold. The resulting channel dependence highlights a potential phenomenological handle for disentangling rotational effects in heavy-ion collisions: while light dilepton spectra encode the imprints of vorticity in the infrared sector, the muon channel provides a comparatively robust baseline. 
\end{abstract}

\maketitle 
\section{Introduction}

As famously stated, \textit{E pur si muove}, attributed to Galileo Galilei~\cite{drake2003galileo}, even small rotations can induce profound physical consequences. Peripheral heavy-ion collisions provide an unparalleled environment for probing strongly interacting matter under extreme conditions. In particular, such collisions generate ultra-intense magnetic fields within the scale of the nuclear overlap region, whose strength increases with the impact parameter and may reach values on the order of the pion mass squared ($\sim 10^{18}$~G)~\cite{PhysRevC.83.054911,BZDAK2012171,CastanoYepes2021}. These conditions enable the study of deconfined QCD degrees of freedom in the presence of strong background fields, which in turn give rise to a wide range of novel effects on particle production rates, transport properties, and symmetry-breaking patterns. For a comprehensive review of strongly interacting matter in intense magnetic backgrounds, see Ref.~\cite{Adhikari_2026,Adhikari2024}. Beyond magnetic fields, the vorticity of the medium formed in non-central collisions, arising from the spatial anisotropy of the matter distribution in the transverse plane, has emerged as another central aspect of current heavy-ion phenomenology~\cite{PhysRevC.77.024906,Becattini2015,PhysRevC.93.064907,PhysRevC.87.034906,PhysRevC.90.024901,Becattini2015}. In such systems, it is believed that a well-defined angular velocity $\mathbf{\Omega}$ is produced, with estimates reaching $\sim 10^{22}~\text{s}^{-1}$ ($\sim 7$ MeV)~\cite{Adamczyk2017}. Unlike the magnetic field, which although extremely strong is short-lived~\cite{PhysRevC.107.034901}, the vorticity is expected to persist throughout the thermal stages of the evolution, particularly within the quark–gluon plasma (QGP). As such, it may decisively shape a variety of observables across both the early and thermalized phases of the collision.  

Recent theoretical efforts have incorporated vorticity as a fundamental dynamical ingredient capable of modifying several observables that were previously analyzed in the absence of rotation. For instance, the restoration of chiral symmetry in QCD has been revisited in rotating backgrounds within effective approaches such as the Yukawa and linear sigma models, where both scalar and fermionic degrees of freedom are influenced by the rotational environment~\cite{PhysRevD.111.036003,PhysRevD.108.094020}. Furthermore, the interplay between magnetic fields and vorticity has attracted significant attention, particularly in relation to their coupled generation mechanisms, as both may arise from the interaction of entropy or charge-density gradients with velocity gradients in the hydrodynamic description of the medium~\cite{PhysRevD.90.083001}. Vorticity is also expected to impact hadron polarization phenomena, including the global polarization of hyperons and vector mesons, which are considered sensitive probes of the rotational structure of the medium~\cite{PhysRevC.76.044901,PhysRevC.84.054910,Wei_2022_ChinesePhysicsC}. These studies underscore the importance of consistently accounting for vorticity in the dynamical modeling of ultrarelativistic nuclear collisions.  

In this work, we focus on dilepton production in a thermalized and rotating QGP. As penetrating electromagnetic probes, dileptons play a central role in current and future experimental programs. Unlike hadrons, whose interactions with the surrounding medium substantially modify their spectral properties, dileptons possess mean free paths that exceed the typical size of the fireball and thus carry essentially undistorted information about the conditions of the plasma from which they originate to the detector~\cite{PhysRevC.92.024912,PhysRevLett.127.042302,PhysRevLett.121.132301}. The dilepton production rate has been systematically studied within a wide range of theoretical frameworks, incorporating the effects of finite temperature~\cite{PhysRevD.107.036017,DUSLING2008246,PhysRevD.93.034017,Martinez_2008,doi:10.1139/cjp-2012-0554}, static external magnetic fields~\cite{PhysRevD.106.056021,PhysRevD.106.036014,PhysRevD.95.074019,PhysRevC.88.024910,Mondal2023}, and even fluctuating magnetic backgrounds~\cite{PhysRevD.111.076028}. Several complementary methods have been used, including the Ritus eigenfunction approach~\cite{SADOOGHI2017218}, real-time thermal field theory~\cite{PhysRevD.98.076006}, and photon flux techniques~\cite{PhysRevD.106.036014}, among others~\cite{PhysRevD.101.096002,PhysRevD.103.096021}.  

To establish the role of vorticity in these processes, we analyze the photon polarization tensor in a thermal medium. {Previous estimations of the photon polarization tensor in the presence of vorticity were reported in Refs.~\cite{Wei_2022_ChinesePhysicsC,Wei_2022_PhysRevD.105.054014}, and more recently extensions involving the presence of a magnetic field as well~\cite{das2023quarkpropagatordileptonproduction}. These computations require analytical expressions for the Fermion propagator in the presence of vorticity, where previous approximations have been presented in~\cite{PhysRevD.103.076021,PhysRevD.104.039901} for open boundary conditions, and an extensive mathematical analysis with different boundary conditions presented in~\cite{Ambrus_PhysRevD.93.104014}. In these previous results and approximations, the non-rotating limit is not smoothly connected with the rotating case. Therefore, we shall use here a different approximation for the Fermion propagator, mainly based on the formalism developed by Vilenkin~\cite{Vilenkin_PhysRevD.21.2260}, as discussed in detail in Appendix~\ref{AppProp}. 

The photon polarization tensor} is central to the description of QCD/QED plasmas, as it governs not only the production of dileptons and photons, but also the collective excitations of the medium. Its evaluation has traditionally included thermal and magnetic effects~\cite{PhysRevD.101.036016,PhysRevD.102.076010,gluon_thermal_ayala,Ayala2021,PhysRevD.109.056008,PhysRevD.109.056007}. Here, we extend this framework to incorporate a vortical background, aiming to identify its imprints on the dilepton emission rate. Such an analysis is timely in light of recent measurements of polarization and flow-sensitive observables at RHIC and the LHC~\cite{Adamczyk2017,PhysRevC.98.014910}, which highlight the need for a systematic theoretical understanding of rotational effects in electromagnetic probes. {Our theoretical analysis of the role of vorticity in the dilepton emission rate at finite temperatures, based on the computation of the photon polarization tensor, complements previous results that reported the dilepton yield and elliptic flow in a rotating QGP~\cite{Wei_2022_PhysRevD.105.054014}, as well as previous estimations of the dilepton helical rate~\cite{Dong_2022}.}

The paper is organized as follows. In Sec.~\ref{Theory}, we present the general formalism connecting the angular-resolved dilepton emission rate with the photon polarization tensor, and we define the coordinate system and approximations assumed for the fermion propagators in a thermal vortical medium. {Details on the derivation of the propagator are presented in Appendix~\ref{AppProp}}. In Sec.~\ref{Polarization_Tensor}, we derive our analytical expressions for the photon polarization tensor at finite temperature and vorticity, with additional details provided in Appendixes~A--D. In Sec.~\ref{Results}, we present our results for the dilepton rates, distinguishing among different fermion–antifermion channels. Finally, in Sec.~\ref{Conclusions}, we summarize our findings and outline their phenomenological implications.

\section{Theory}
\label{Theory}

\begin{figure}[h!]
    \centering
    \includegraphics[scale=0.15]{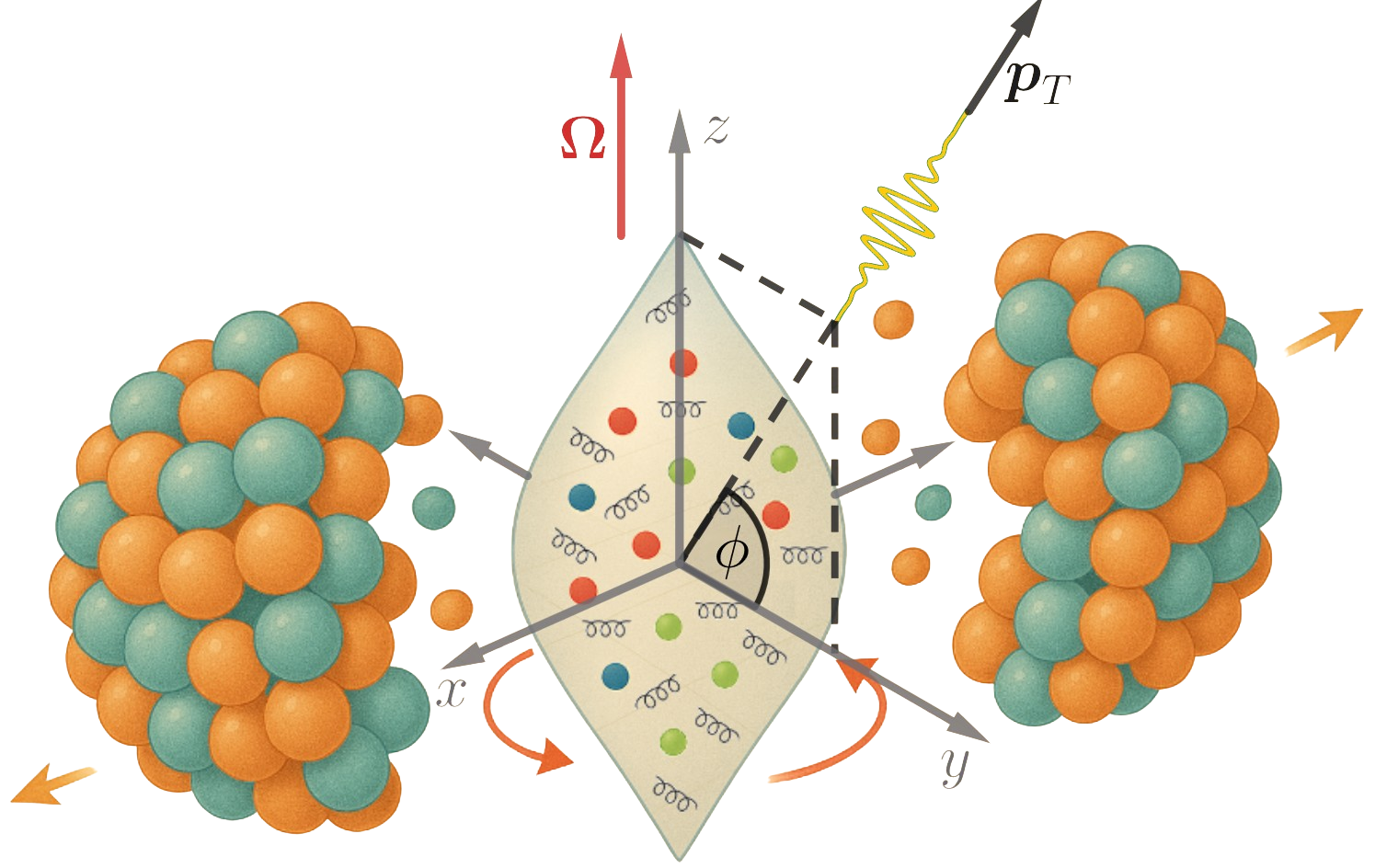}
    \caption{Coordinate system for the photon's momentum and the local vorticity in the nuclear collision region. Schematic generated using an AI model and Geogebra~\cite{openai_chatgpt,geogebra}.}
    \label{fig:coordSystemOmega}
\end{figure}

\begin{figure}[h!]
    \centering
    \includegraphics[scale=0.6]{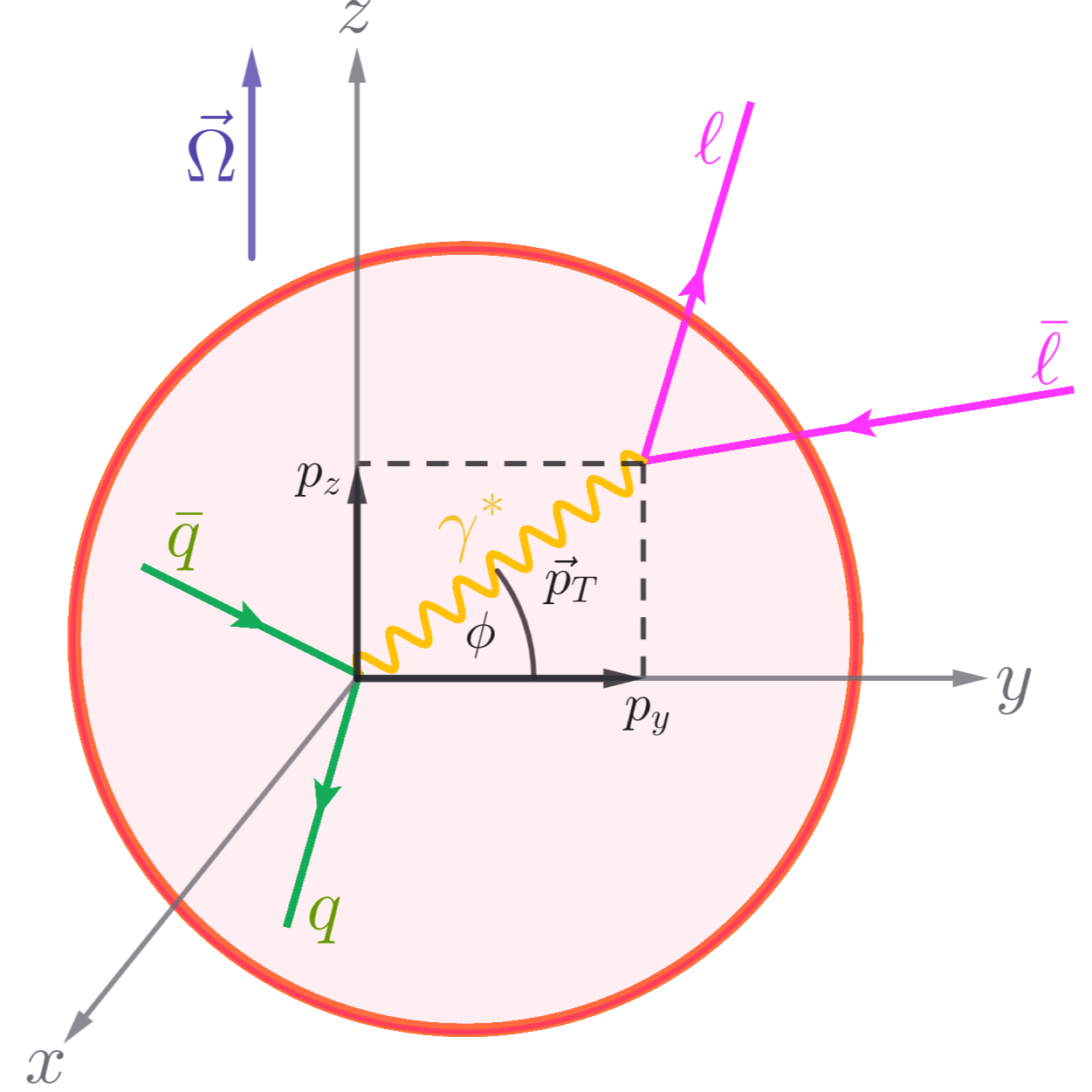}
    \caption{Feynman diagram and coordinates system for the process of dilepton emission by quark-antiquark annihilation mediated by a virtual photon $q \overline{q}\rightarrow\gamma\rightarrow l\overline{l}$.}
    \label{fig:Feynman}
\end{figure}

The primary objective of this work is to compute the dilepton emission rate distribution, assuming that the primary nuclear collision generates a local vorticity $\mathbf{\Omega} = \hat{z}\Omega$, as depicted in Fig.~\ref{fig:coordSystemOmega}. We thus consider the process represented in Fig.~\ref{fig:Feynman}, where a quark/anti-quark annihilation is followed by a highly energetic virtual photon decaying into a lepton-anti-lepton pair. Following similar arguments as those discussed in detail in~\cite{PhysRevD.106.036014}, the rate arising from this process is given by the expression
{\small{\bea\label{eq:Rate1}
\mathcal{R}_{\ell\bar{\ell}}&\equiv&\frac{d^4 R_{\ell\bar{\ell}}}{dp^4}\\
&=&\frac{\alpha_\text{em}n_\text{B}(\omega)}{12\pi^4M^2}\left(1+\frac{2m_\ell^2}{M^2}\right)\left(1-\frac{4m_\ell^2}{M^2}\right)^{\frac{1}{2}}\text{Im}\left[g_{\mu\nu}\Pi_\text{R}^{\mu\nu}(\omega)\right],\nn
\eea}}
when $\omega \to \sqrt{M^2+P_T^2}$. Here, $n_\text{B}(\omega) = \left(e^{\beta\omega}-1\right)^{-1}$ denotes the Bose--Einstein distribution at finite temperature, $M$ is the invariant dilepton mass, and $\Pi_\text{R}^{\mu\nu}(\omega)$ represents the retarded photon polarization tensor that includes all the quark/anti-quark possible channels~\cite{PhysRevD.106.036014}. The geometry shown in Fig.~\ref{fig:Feynman} defines the transverse momentum as
\bea
p_T &=& \sqrt{p_y^2 + p_z^2},\nn\\
p_y &=& p_T \cos\phi,\nn\\
p_z &=& p_T \sin\phi.
\label{eq:coordinates}
\eea

The longitudinal momentum is expressed in terms of the rapidity
\bea
y = \frac{1}{2}\ln\left(\frac{p_0 + p_x}{p_0 - p_x}\right),
\eea
and, in particular, we focus on the case of midrapidity, $y=0$ (corresponding to $p_x = 0$).

\begin{figure}[h!]
    \centering
    \includegraphics[scale=0.4]
    {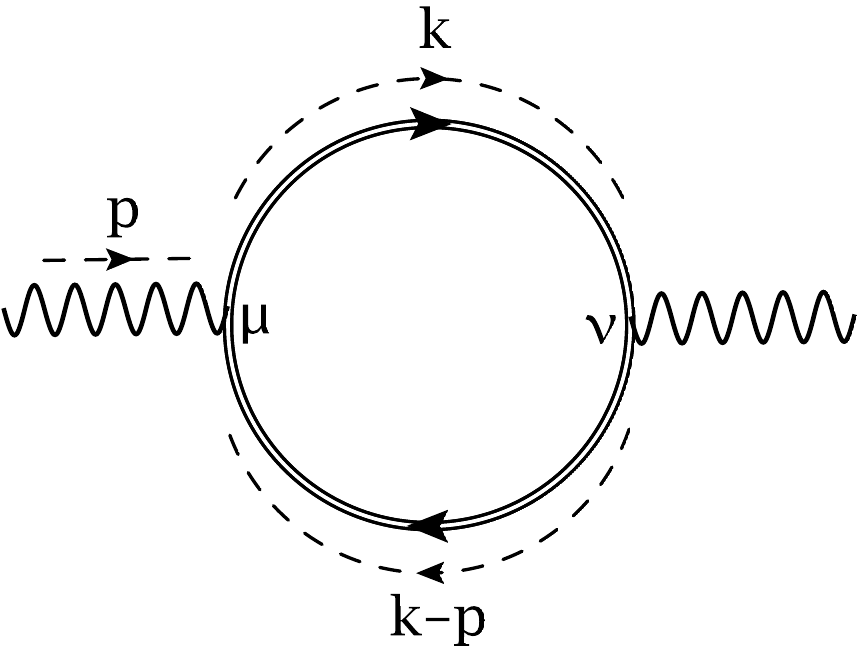}
    \caption{Feynman diagram illustrating the the polarization tensor. The dashed arrows indicate the charge flow in the diagram.}
    \label{fig:pol1}
\end{figure}

In the one-loop approximation, the photon polarization tensor due to a virtual quark/anti-quark pair of charge $q_f$, is represented by the Feynman diagram shown in Fig.~\ref{fig:pol1}, whose corresponding mathematical expression is given by:
{\small
\bea
\ii\Pi^{\mu\nu}=-\frac{1}{2}\int\frac{d^4k}{(2\pi)^4}\text{Tr}\Big\{\ii q_f\gn\ii S\left(k\right)\ii q_f\gm\ii S(k-p)\Big\}.
\label{eq:PolTR}
\eea}
{
\subsection{The Fermion propagator in a rotating frame}

We further assume a rotating environment with a constant angular velocity $\mathbf{\Omega}$. The relative tangential velocity of the co-rotating frame with respect to the laboratory frame is $\mathbf{V} = \mathbf{\Omega}\times\mathbf{R}$. For a choice of the rotation axis such that $\mathbf{\Omega} = \hat{z}\Omega$, as depicted in Fig.~\ref{fig:coordSystemOmega}, the co-rotating frame is related to the laboratory frame by a system of tetrads (vierbeins)~\cite{Weinberg} $\left\{ e^{a}_{\,\mu} \right\}$. Here, $\mu = \hat{0},\, \hat{1},\,\hat{2},\, \hat{3}$ are Minkowski indexes in the laboratory frame, while $a = 0,\,1,\,2,\,3$ are those in the (locally flat) co-rotating frame. Therefore, we explicitly have
\begin{eqnarray}
e_{0}^{\,\hat{0}}= e_{1}^{\,\hat{1}} = e_{2}^{\,\hat{2}} = e_{3}^{\,\hat{3}} = 1, \,\,\,e_{0}^{\,\hat{1}} = \Omega y,\,\,e_{0}^{\,\hat{2}} =  -\Omega x.
\end{eqnarray}
The corresponding local metric $g_{\mu\nu} = \eta_{ab}e^{a}_{\,\mu}e^{b}_{\,\nu}$ is given by
\begin{eqnarray}
g_{\mu\nu} = \left( \begin{array}{cccc} 1 - \Omega^2\left( x^2 + y^2 \right) & \Omega y & -\Omega x & 0\\\Omega y & -1 & 0 & 0\\-\Omega x & 0 & -1 & 0\\0 & 0 & 0 & -1\end{array}\right)
\end{eqnarray}
The Dirac matrices in the laboratory frame are defined, in terms of the vierbein, by~\cite{Birrell} $\hat{\gamma}^{\mu}(x) = e_{a}^{\,\mu}\gamma^{a}$, 
so that while the $\gamma^{a}$ in the local co-rotating frame satisfy the usual Clifford algebra with the flat Minkowski metric $\left[\gamma^{a},\gamma^b \right] = 2 \eta^{ab}$, those in the laboratory frame satisfy~\cite{Birrell}  $\left[ \hat{\gamma}^{\mu},\hat{\gamma}^{\nu}\right] = 2 g^{\mu\nu}$.

The Fermion propagator is constructed, as explained in detail in Appendix~\ref{AppProp}, from the free Dirac Lagrangian in the laboratory frame,
\bea
\mathcal{L} = \overline{\psi}\left[ \ii\hat{\gamma}^{\mu}D_{\mu} - m_f \right]\psi.
\eea
Here, in the definition of the covariant derivative $D_{\mu}\equiv \partial_{\mu} + \Gamma_{\mu}$ we included the spin connection~\cite{Birrell,Weinberg}
\bea
\Gamma^{\mu} = -\frac{\ii}{4}\omega^{\mu}_{\,ab}\sigma^{ab},
\eea
for $\sigma^{ab} = \left( \ii/2  \right)\left[ \gamma^a,\gamma^b \right]$, and
\bea
\omega_{\mu\,ab} \equiv g_{\lambda\rho}e^{a}_{\,\lambda}\left(  
\partial_{\mu}e_{b}^{\,\rho} + \Gamma_{\mu\nu}^{\rho}e_b^{\,\nu}
\right), 
\eea
defined after the Christoffel symbols~\cite{Weinberg}
\bea
\Gamma_{\mu\nu}^{\lambda} = \frac{1}{2}g^{\lambda\sigma}\left(  
\partial_{\mu} g_{\sigma\nu} + \partial_{\nu} g_{\mu\sigma} - \partial_{\sigma} g_{\mu\nu}
\right).
\eea
As shown in detail in Appendix~\ref{AppProp}, the only non-vanishing component of the spin connection is $\Gamma_{\hat{0}} = -\left(\ii/2\right)\Omega\sigma^{\hat{1}\hat{2}} = -\left(\ii/2\right)\mathbf{\Omega}\cdot{\boldsymbol{\Sigma}}$, proportional to the z-component of the spin operator $\boldsymbol{\Sigma}$. The corresponding form of the free Lagrangian, including the finite chemical potential $\mu$, is
\bea
\mathcal{L} = \overline{\psi}\left[ \ii\gamma^a\partial_a + \gamma^0\left( \Omega J_z + \mu  \right) - m_f  \right]\psi(x), 
\eea
with $\mathbf{J} = \mathbf{L} + \frac{1}{2}\mathbf{\Sigma}$ the total angular momentum, involving orbital $\mathbf{L}$ and spin $\frac{1}{2}\boldsymbol{\Sigma}$, respectively.
}

{Therefore, the Fermion propagator is described accordingly, where in the {\it{rigid rotation}} approximation developed by us from the Vilenkin formalism~\cite{Vilenkin_PhysRevD.21.2260}, as shown in Appendix~\ref{AppProp}, it reads:}
\bea
S(p)&=&\Op{+}\frac{\slashed{p}_++m_f}{p_+^2-m_f^2+\ii\epsilon}+\Op{-}\frac{\slashed{p}_-+m_f}{p_-^2-m_f^2+\ii\epsilon}
\eea

Here, we defined the spin projectors along the direction of the angular velocity, as follows
\begin{subequations}
    \bea
    \Op{\pm}= \frac{1}{2}\left(\mathbb{1}\pm \frac{\mathbf{\Omega}}{\Omega}\cdot\mathbf{\Sigma}\right) = \frac{1}{2}\left(\mathbb{1}\pm\ii\gamma^1\gamma^2\right)
\eea
and
\bea
p^\mu_\pm\equiv\left(p^0\pm\frac{\Omega}{2},\mathbf{p}\right).
\eea
\end{subequations}

We remark that, in the rigid rotating approximation, the angular velocity $\mathbf{\Omega}$ plays a role analogous to a spin-dependent chemical potential, and hence it does not break the $SO(3)$ rotational symmetry of the Fermion propagators.

In order to obtain the retarded polarization tensor as required by the definition of the emission rate Eq.~\eqref{eq:Rate1}, we first compute the diagrams in Matsubara space, and then we perform the standard analytic continuation onto real frequency according to the prescription
\be
\Pi_R^{\mu\nu}(p^0 = \omega,\mathbf{p}) = \Pi^{\mu\nu}(\ii\nu_n\rightarrow \omega + \ii\epsilon,\mathbf{p}),
\ee
where we finally set the external frequency $\omega \to \sqrt{M^2+p_T^2}$, for $M$ the invariant dilepton mass.
\section{The Photon Polarization Tensor}
\label{Polarization_Tensor}




As shown in Appendix~\ref{Ap:ComputationOfPi}, the retarded photon polarization tensor for a quark/anti-quark pair with charge $q_f$ and mass $m_f$ takes the form
{
\bea
g_{\mu\nu}\Pi^{\mu\nu}_R(\omega) &=& 
\frac{q_f^2}{8\pi}\Bigg[ 2 m_f^2 \mathcal{F}\left( \beta,\omega,\Omega \right)\nn\\
&&+ \left[(\omega + \sigma\Omega)^2 - \wph^2 \right]\mathcal{G}\left( \beta,\omega,\Omega \right)\Bigg]
\label{eq:PiRcontracted}
\eea
where we defined the functions
\bea
\mathcal{F}\left( \beta,\omega,\Omega \right) &=& \sum_{\sigma,s=\pm 1} \int_{0}^{\infty}dE f_s(E,\omega)\Delta n_F(E,\omega,\sigma\Omega)\nn\\
\label{eq_FF}
\eea
and
\bea
\mathcal{G}\left( \beta,\omega,\Omega \right) &=& \sum_{\sigma,s=\pm 1} \int_{0}^{\infty}dE f_s(E,\omega+\sigma\Omega)\Delta n_F(E,\omega,\sigma\Omega)\nn\\
\label{eq_GG}
\eea
respectively.
In both expressions above, we defined the functions
\bea
\Delta n_{F}\left( E,\wph,\sigma\Omega\right) &=& \omega^{-1}\Bigg\{ n_\text{F}\left[\beta\left( E-\frac{\sigma\Omega}{2}\right)\right]\nn\\
&-& n_\text{F}\left[\beta\left( E - \wph - \frac{\sigma\Omega}{2}\right)\right]\Bigg\}.
\eea
and
\bea
f_s(E,\omega) &=& \Theta(E - m_f)\cdot
\Bigg\{ 
\Theta(E - s\omega - m_f)\nn\\ 
&&- \Theta(-E+s\omega-m_f)\Bigg\}.
\label{eq_fs}
\eea
}
Here, $n_\text{F}(x)=(e^x+1)^{-1}$ is the Fermi--Dirac distribution, and $\Theta(x)$ is the Heaviside step function. {Explicit analytical formulae are obtained for $\mathcal{F}$ (Eq.~\eqref{eq_AFfinal}) and $\mathcal{G}$ (Eq.~\eqref{eq_AGfinal}), respectively, but they are quite long so we present them in the Appendix~\ref{AppD}. We also show, in Appendix~\ref{AppE}, that in the limit $\Omega \rightarrow 0$, our analytical expression for the rate reduces to the formula reported by Weldon~\cite{Weldon_98_PhysRevD.28.2007}.}

Mathematical details, involving evaluation of the sums in Matsubara space and subsequent analytical continuation back to Minkowski space are presented in Appendices~\ref{Ap:MatsubaraSums} and~\ref{Ap.momentumIntegrals}. 

It is worth emphasizing that the effect of vorticity manifests itself as an effective spin-dependent chemical-potential shift, since it only modifies the argument of the Fermi--Dirac distributions.

\section{Results}
\label{Results}

\begin{figure*}
    \centering
    \includegraphics[scale=0.73]{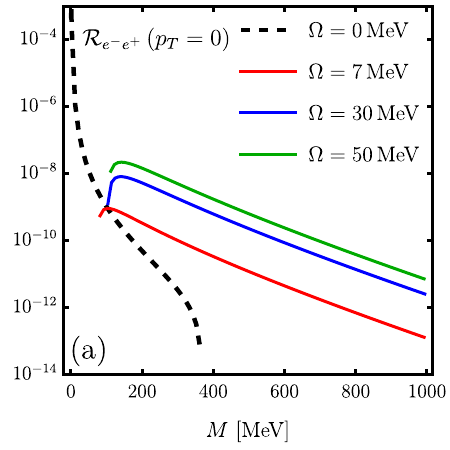}~~~~~\includegraphics[scale=0.73]{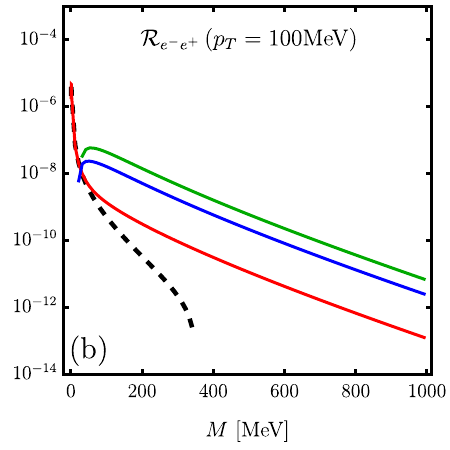}~~~~~\includegraphics[scale=0.73]{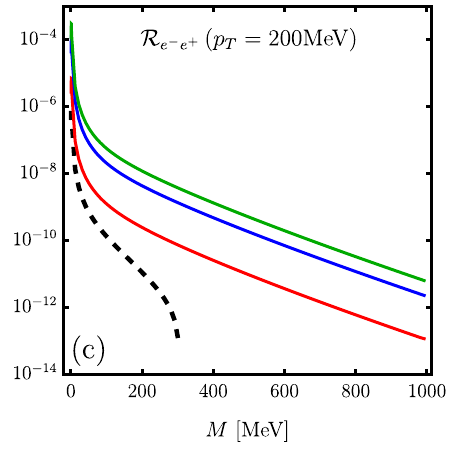}
    \caption{Electron–positron production rate from Eq.~\eqref{eq:Rate1} for several values of the vorticity $\Omega$ (dashed line $\Omega = 0$, red $\Omega = 7$~MeV, $\Omega = 30$~MeV, $\omega = 50$~MeV) and transverse momentum $p_T$. The temperature is fixed at $T = 150$ MeV.}
    \label{fig:ee}
\end{figure*}

\begin{figure*}
    \centering
    \includegraphics[scale=0.73]{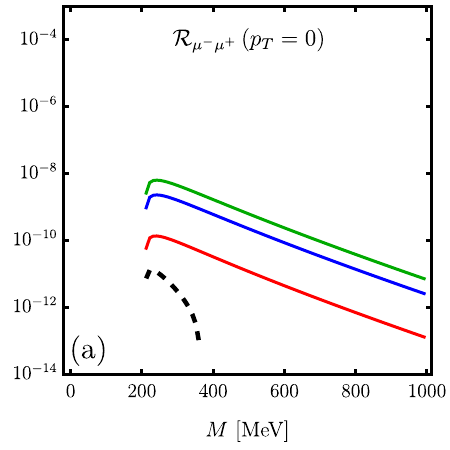}~~~~~\includegraphics[scale=0.73]{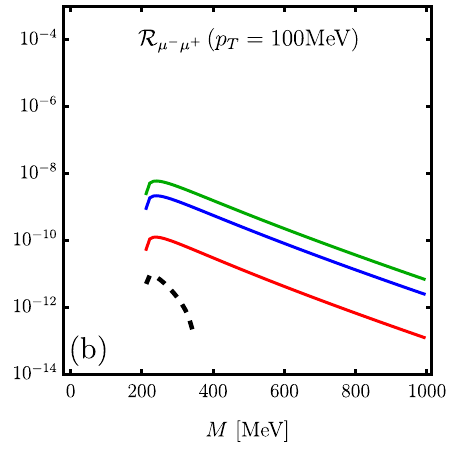}~~~~~\includegraphics[scale=0.73]{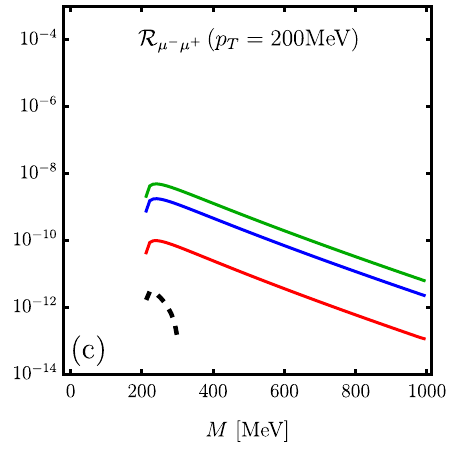}
    \caption{Muon–anti-muon production rates from Eq.~\eqref{eq:Rate1} for different values of the vorticity $\Omega$ (dashed line $\Omega = 0$, red $\Omega = 7$~MeV, $\Omega = 30$~MeV, $\omega = 50$~MeV) and transverse momentum $p_T$. The temperature is set to $T = 150$ MeV. }
    \label{fig:mm}
\end{figure*}

In {Fig.~\ref{fig:ee} and Fig.~\ref{fig:mm}} we show the scaled dilepton production rate, defined as
\bea
\widetilde{\mathcal{R}}_{\ell\ell}\equiv\frac{12\pi^4 M^2}{\alpha_\text{em}}\mathcal{R}_{\ell\ell},
\label{eq:ScaledRate}
\eea
plotted as a function of the transverse energy scale $\sqrt{M^2 + p_T^2}$ for several values of vorticity $\Omega$, at a fixed temperature $T=150~\text{MeV}$. The two figures correspond to different production channels:  $e^-e^+$ pairs {[Fig.~\ref{fig:ee}]}, and $\mu^-\mu^+$ pairs {[Fig.~\ref{fig:mm}]}. In each case, the result for $\Omega=0$, {as obtained from our expressions that agree with Ref.~\cite{Weldon_98_PhysRevD.28.2007} in this limit (see Appendix~\ref{AppE})} is included for comparison. The values of $\Omega$ considered extend up to $50$~MeV, which is consistent with estimates of the maximum vorticity attainable in heavy-ion collisions without violating causality~\cite{PhysRevC.92.014906,PhysRevC.95.054915}. We thus explore this upper bound as a reference scale for comparison.  

For vanishing vorticity, the spectra follow the expected thermal behavior {(see Appendix~\ref{AppE} and Ref.~\cite{Weldon_98_PhysRevD.28.2007})}, showing a monotonically decreasing dependence on the transverse energy scale. However, once rotational effects are introduced, clear modifications appear in the light channel $e^-e^+$, where it is observed that finite values of $\Omega$ suppress the dilepton yield in the low-energy region, with the suppression effect becoming monotonically stronger as $\Omega$ increases from $7$ to $50$~MeV. In addition, we clearly identify the presence of a threshold in dilepton production originating from the Heaviside step functions in Eqs.~(\ref{eq_FF}), (\ref{eq_GG}) as defined in Eq.~\eqref{eq_fs}.  

In the present case, however, the position of this threshold is slightly modified by the vorticity factor, which effectively shifts the argument of the Fermi--Dirac distributions. From a physical standpoint, {in the context of the {\it{rigid rotation approximation}} adopted in our present analysis}, vorticity plays the role of an effective spin-dependent chemical potential associated with the rotational background, thereby redistributing the available phase space for quark--antiquark annihilation. This interpretation naturally explains both the small displacement of the production threshold and the overall suppression of the emission rate at low $\sqrt{M^2 + p_T^2}$. At larger transverse energies the spectra asymptotically converge to the $\Omega=0$ limit, indicating that the impact of vorticity is mostly confined to the infrared sector. Such behavior is in line with previous analyzes of medium-induced modifications to dilepton rates, where the dominant effects also manifest themselves at low invariant masses.  


The behavior of the $\mu^-\mu^+$ channel, shown in Fig.~\ref{fig:mm}, is qualitatively different. Here, the production rate is dominated by the larger mass threshold, which substantially suppresses the spectrum already in the absence of vorticity. As a result, the relative effect of the rotational background is {milder for muon-anti-muon pairs than for lighter electron-positron pairs}. This { example illustrates} that heavier dilepton channels are less affected by modifications of the thermal distribution {due to vorticity}, thus providing a more stable probe of the medium.  

Together, these results highlight an important channel dependence in the response of dilepton production to vorticity. Light dilepton channels, namely the electron-positron channel, display sizable modifications in the low-mass region, while the heavier muon channel {a milder response}. From a phenomenological perspective, this difference suggests that experimental measurements of dilepton spectra in heavy-ion collisions could in principle disentangle rotational effects by comparing light and heavy dilepton yields. In particular, any suppression in the low-energy sector of the electron channel relative to the muon channel may serve as an indicator of vorticity in the medium created during the early stages of the collision.

\section{Summary and Conclusions}\label{Conclusions}

In this work we have analyzed dilepton production in a thermalized and rotating quark--gluon plasma by computing the photon polarization tensor at finite temperature and vorticity. This framework allowed us to quantify the role of global rotation in shaping the dilepton emission rate in two different channels, thereby {complementing previous studies based on different approximations to the Fermion propagator in a rotating frame~\cite{Wei_2022_ChinesePhysicsC,Wei_2022_PhysRevD.105.054014}}.

Our main findings can be summarized as follows: For the $e^-e^+$ channel, finite values of the angular velocity $\Omega$ lead to a clear suppression of the dilepton yield in the low-mass region, together with a mild displacement of the pair production threshold. These effects can be traced back to the modification of the Fermi distribution functions, as well as the pair production threshold, in the presence of vorticity which, {in the rigid rotating approximation, acts as an effective spin-dependent rotational chemical potential} redistributing the phase space available for the process. The impact of rotation is more appreciable in the infrared sector of the spectrum, while at larger transverse energy scales (and invariant dilepton masses) the rates converge to their baseline thermal behavior { at $\Omega = 0$}. By contrast, the heavier $\mu^-\mu^+$ channel shows {a milder} sensitivity to vorticity, as its production rate is already strongly suppressed by the larger intrinsic mass threshold {at $\Omega = 0$}.

Together, these results point to a pronounced mass dependence in the response of dilepton production to rotation. From a phenomenological standpoint, this opens an avenue for using comparative analyses of light- and heavy- dilepton yields as a probe of vorticity in ultrarelativistic nuclear collisions. In particular, a relative suppression of the electron channel in the low-mass region with respect to the muon channel could serve as a signature of rotational effects in the quark--gluon plasma. Such a strategy would complement ongoing efforts to extract information on vorticity from hadronic polarization measurements, while providing an independent electromagnetic probe that is less affected by interactions in the final-state. 

Looking ahead, we are extending the present analysis {going beyond the rigid rotating approximation, by incorporating the effects of the higher angular momentum channels. Moreover,}
additional ingredients such as time-dependent vorticity profiles, realistic collective flow dynamics, and the interplay with strong magnetic fields are under consideration. We are also investigating differential dilepton observables in rapidity and azimuthal angle, which would allow for a more direct comparison with experimental measurements. We are actively seeking these avenues, and this is work in progress that will be presented elsewhere.

{
\section*{Acknowledgments}

JDCY would like to thank the Isaac Newton Institute for Mathematical Sciences, Cambridge, for support and hospitality during the programme \textit{Quantum field theory with boundaries, impurities, and defects (BID2025)} where work on this paper was undertaken. This work was supported by EPSRC grant no.\ EP/R014604/1. EM acknowledges ANID Fondecyt 
grant no. 1230440 that supported this work.
}

%

\appendix
\onecolumngrid

{
\section{The Fermion propagator}
\label{AppProp}
As stated in the main text, we assume a rotating frame with a constant angular velocity $\mathbf{\Omega}$. The relative tangential velocity of the co-rotating frame relative to the laboratory frame is $\mathbf{V} = \mathbf{\Omega}\times\mathbf{R}$. For a choice of the rotation axis such that $\mathbf{\Omega} = \hat{z}\Omega$, as depicted in Fig.~\ref{fig:coordSystemOmega}, the co-rotating frame is related to the laboratory frame by a system of tetrads (vierbeins)~\cite{Weinberg} $\left\{ e^{a}_{\,\mu} \right\}$. Here, $\mu = \hat{0},\, \hat{1},\,\hat{2},\, \hat{3}$ are Minkowski indexes in the laboratory frame, while $a = 0,\,1,\,2,\,3$ are those in the (locally flat) co-rotating frame. Therefore, we explicitly have~\cite{Vilenkin_PhysRevD.21.2260,Ambrus_PhysRevD.93.104014,PhysRevD.103.076021}
\begin{eqnarray}
e_{0}^{\,\hat{0}}= e_{1}^{\,\hat{1}} = e_{2}^{\,\hat{2}} = e_{3}^{\,\hat{3}} = 1, \,\,\,e_{0}^{\,\hat{1}} = \Omega y,\,\,e_{0}^{\,\hat{2}} =  -\Omega x.
\label{eq_vierbein}
\end{eqnarray}
The corresponding local metric $g_{\mu\nu} = \eta_{ab}e^{a}_{\,\mu}e^{b}_{\,\nu}$ is given by
\begin{eqnarray}
g_{\mu\nu} = \left( \begin{array}{cccc} 1 - \Omega^2\left( x^2 + y^2 \right) & \Omega y & -\Omega x & 0\\\Omega y & -1 & 0 & 0\\-\Omega x & 0 & -1 & 0\\0 & 0 & 0 & -1\end{array}\right).
\label{eq_gmunu}
\end{eqnarray}
The Dirac matrices in the laboratory frame are defined, in terms of the vierbein, by~\cite{Birrell} $\hat{\gamma}^{\mu}(x) = e_{a}^{\,\mu}\gamma^{a}$, 
so that while the $\gamma^{a}$ in the local co-rotating frame satisfy the usual Clifford algebra with the flat Minkowski metric~\cite{Birrell} $\left[\gamma^{a},\gamma^b \right] = 2 \eta^{ab}$, those in the laboratory frame satisfy $\left[ \hat{\gamma}^{\mu},\hat{\gamma}^{\nu}\right] = 2 g^{\mu\nu}$. The explicit expressions for these matrices are
\bea
\hat{\gamma}^{\hat{0}} = \gamma^0,\,\,\hat{\gamma}^{\hat{1}} = \Omega y \gamma^{0} + \gamma^1,\,\,\gamma^{\hat{2}} = -\Omega x \gamma^{0} + \gamma^2,\,\,\gamma^{\hat{3}} = \gamma^3.
\label{eq_gammamu}
\eea
The Fermion propagator is constructed from the free Dirac Lagrangian in the laboratory frame,
\bea
\mathcal{L} = \overline{\psi}\left[ \ii\hat{\gamma}^{\mu}D_{\mu} - m_f \right]\psi.
\eea
Here, in the definition of the covariant derivative $D_{\mu}\equiv \partial_{\mu} + \Gamma_{\mu}$ we included the spin connection~\cite{Weinberg,Birrell}
\bea
\Gamma^{\mu} = -\frac{\ii}{4}\omega^{\mu}_{\,ab}\sigma^{ab},
\label{eq_Connection}
\eea
for $\sigma^{ab} = \left( \ii/2  \right)\left[ \gamma^a,\gamma^b \right]$, and
\bea
\omega_{\mu\,ab} \equiv g_{\lambda\rho}e^{a}_{\,\lambda}\left(  
\partial_{\mu}e_{b}^{\,\rho} + \Gamma_{\mu\nu}^{\rho}e_b^{\,\nu}
\right),
\label{eq_omegamunu}
\eea
defined after the Christoffel symbols~\cite{Weinberg}
\bea
\Gamma_{\mu\nu}^{\lambda} = \frac{1}{2}g^{\lambda\sigma}\left(  
\partial_{\mu} g_{\sigma\nu} + \partial_{\nu} g_{\mu\sigma} - \partial_{\sigma} g_{\mu\nu}
\right).
\label{eq_Christoffel}
\eea
The explicit analytical expressions for the non-vanishing Christoffel symbols, from Eq.~\eqref{eq_gmunu}, are
\bea
\Gamma_{\hat{0}\hat{2}}^{\hat{1}} = \Gamma_{\hat{2}\hat{0}}^{\hat{1}} = - \Omega,\,\,\Gamma_{\hat{0}\hat{1}}^{\hat{2}} = \Gamma_{\hat{1}\hat{0}}^{\hat{2}} = \Omega,\,\,\Gamma_{\hat{0}\hat{0}}^{\hat{1}} = - \Omega^2\,x,\,\,\Gamma_{\hat{0}\hat{0}}^{\hat{2}} = -\Omega^2\,y.
\eea
Combining these expressions with the vierbein Eq.~\eqref{eq_vierbein}, after Eq.~\eqref{eq_omegamunu} we obtain the only non-vanishing component of the spin connection, $\Gamma_{\hat{0}} = -\left(\ii/2\right)\Omega\sigma^{\hat{1}\hat{2}} = -\left(\ii/2\right)\mathbf{\Omega}\cdot{\boldsymbol{\Sigma}}$, which is proportional to the z-component of the spin operator $\boldsymbol{\Sigma}$. The corresponding form of the free Lagrangian density is
\bea
\mathcal{L} = \overline{\psi}\left[ \ii\gamma^a\partial_a + \gamma^0\Omega \hat{J}_z  - m_f  \right]\psi(x), 
\label{eq_ApLagrange}
\eea
with $\hat{\mathbf{J}} = \hat{\mathbf{L}} + \frac{1}{2}\hat{\mathbf{\Sigma}}$ the total angular momentum, involving orbital $\mathbf{L}$ and spin $\frac{1}{2}\boldsymbol{\Sigma}$, respectively. For this Fermion Lagrangian density, the corresponding Hamiltonian commutes with the set of three operators $\left\{ \hat{P}_z, \hat{J}_z, \hat{W}_0 \right\}$, where $\hat{P}_z$ is the z-component of the linear momentum operator $\hat{\mathbf{P}}$, $\hat{J}_z = \hat{L}_z + \frac{1}{2}\hat{\Sigma}_z = -\ii\partial_{\varphi} + \frac{1}{2}\hat{\Sigma}_z$ the z-component of the total angular momentum operator, and $\hat{W}_0$ the 0-component of the Pauli-Lubanski operator~\cite{Itzykson},
\bea
\hat{W}_0 = \frac{\hat{\mathbf{P}}\cdot\hat{\mathbf{J}}}{P},
\eea
whose eigenvalues $h = \pm 1$ represent the helicity components of the spinor eigenfunctions.

The fermion propagator is expressed in terms of the eigenbasis spanned by the solutions to the Dirac equation in rotating coordinates, that after the Lagrangian density Eq.~\eqref{eq_ApLagrange} is
\bea
\left[ \gamma^0\left( \ii\partial_t + \Omega\hat{J}_z + \ii\vec{\gamma}\cdot\nabla - m_f \right) \right]\hat{\psi}(x) = 0.
\eea
The corresponding spinor solutions, for positive ($\hat{U}$) and negative ($\hat{V}$) energy eigenvalues are given by~\cite{Ambrus_PhysRevD.93.104014}
\bea
\hat{U}_{\alpha_{\ell}}(x) &=& \frac{1}{2\pi}e^{-\ii \tilde{E}_j t}  \hat{u}_{\alpha_{\ell}}(\mathbf{x})\nn\\
\hat{V}_{\alpha_{\ell}}(x) &=& \ii\gamma^2 \hat{U}^{*}_{\alpha_{\ell}}(x) \equiv \frac{1}{2\pi}e^{\ii \tilde{E}_j t} \hat{v}_{\alpha_{\ell}}(\mathbf{x}),
\eea
where we defined $\tilde{E}_j \equiv E - \left( \ell + 1/2\right)\Omega$, for $j = \ell + 1/2$ the eigenvalues of $\hat{J}_z$, with $\ell\in\mathbb{Z}$, and the multi-index  
\bea
\alpha_{\ell} \equiv (E,p_z,\ell,h)
\eea
These eigenfunctions are defined in polar coordinates $\mathbf{x} = \left( \rho,\varphi,z \right)$, as follows~\cite{Ambrus_PhysRevD.93.104014}
\bea
\hat{u}_{\alpha_{\ell}}(\mathbf{x}) = e^{\ii p_z z} \left( \begin{array}{c} \C_{\alpha_{\ell}+} \phi_{\alpha_{\ell}} (\rho,\varphi)\\ \C_{\alpha_{\ell}-} \phi_{\alpha_{\ell}}(\rho,\varphi) \end{array}\right),
\eea
with $\C_{\alpha_{\ell}\pm}$ normalization coefficients independent of $\Omega$, and the bi-spinors $\phi_{\alpha_{\ell}}(\rho,\phi)$ satisfying the differential equation~\cite{Ambrus_PhysRevD.93.104014}
\bea
\left( \begin{array}{cc} -\ii\partial_{\varphi} + \frac{1}{2} & 0\\0 & -\ii\partial_{\varphi}-\frac{1}{2} \end{array}\right)\phi_{\alpha_{\ell}}(\rho,\phi) = \left( \ell + \frac{1}{2} \right)\phi_{\alpha_{\ell}}(\rho,\phi),
\eea
so that the spinor solutions are, by definition, also eigenfunctions of the $z$-component of the total angular momentum operator $\hat{J}_z$ (for $j=\ell+1/2$)
\bea
\hat{J}_z\hat{u}_{\alpha_{\ell}}(\mathbf{x}_1) &=& j \hat{u}_{\alpha_{\ell}}(\mathbf{x}_1)\nn\\
\hat{J}_z\hat{v}_{\alpha_{\ell}}(\mathbf{x}_1) &=& -j \hat{v}_{\alpha_{\ell}}(\mathbf{x}_1).
\label{eq_ApJeigen}
\eea
The explicit form of these eigenspinors, however not essential for the analysis of the propagator that follows, is constructed in terms of Bessel functions $J_{\ell}(p_{\perp}\rho)$~\cite{Ambrus_PhysRevD.93.104014}
\bea
\phi_{\alpha_{\ell}}(\rho,\varphi) = \frac{1}{\sqrt{2}}\left( \begin{array}{c} \mathcal{N}_{h} e^{\ii\ell\varphi}J_{\ell}(p_{\perp}\rho)\\ 2\ii h \mathcal{N}_{-h} e^{\ii(\ell + 1)\varphi} J_{\ell + 1}(p_{\perp}\rho)  \end{array}\right),
\eea
with the $\Omega$-independent constants $\mathcal{N}_{h} = \left( 1 + 2 h p_{z}/p\right)^{1/2}$, with $p = \sqrt{p_z^2 + p_{\perp}^2}$ the total linear momentum.

The free-particle propagator, in a rotating system at finite temperature $T = \beta^{-1}$, is expressed by~\cite{Vilenkin_PhysRevD.21.2260}
\bea
S'(\mathbf{x}_1,\tau_1;\mathbf{x}_2,\tau_2) = \beta^{-1}\sum_n e^{-\ii\omega_n (\tau_1-\tau_2)} S'(\mathbf{x}_1,\mathbf{x}_2,\ii\omega_n),
\eea
where (in terms of $a,b$ Dirac-spinor components) we have the spectral decomposition~\cite{Vilenkin_PhysRevD.21.2260,Ambrus_PhysRevD.93.104014}
\bea
S'_{ab}(\mathbf{x}_1,\mathbf{x}_2,\ii\omega_n) = -\sum_{\ell = -\infty}^{\infty}\sum_{h=\pm}\int_{m_f}^{\infty} dE \int_{-p}^{p}dp_z
\left( \frac{u_{\alpha_{\ell}}^{a}(\mathbf{x}_1)\overline{u}^{b}_{\alpha_{\ell}}(\mathbf{x}_2)}{\ii\omega_n - E + \left(\ell + \frac{1}{2}\right)\Omega} + \frac{v^{a}_{\alpha_{\ell}}(\mathbf{x}_1)\overline{v}^{b}_{\alpha_{\ell}}(\mathbf{x}_2)}{\ii \omega_n + E - \left( \ell + \frac{1}{2} \right)\Omega} \right).
\label{eq_Apspectral}
\eea
Here, $\bar{u}_{\alpha_{\ell}}(\mathbf{x}) \equiv \hat{u}^{\dagger}_{\alpha_{\ell}}(\mathbf{x})\gamma^{0}$ and $\bar{v}_{\alpha_{\ell}}(\mathbf{x}) \equiv \hat{v}^{\dagger}_{\alpha_{\ell}}(\mathbf{x})\gamma^{0}$, in the usual notation. 

Let us now consider the identity (for $j = \ell + 1/2$)
\bea
\frac{1}{\ii \omega_n \mp E \pm j\Omega} = \frac{1}{\ii\omega_n \mp E}\left( 1 \pm \frac{j\Omega}{\ii\omega_n \mp E} \right)^{-1} = \frac{1}{\ii\omega_n \mp E}\sum_{k=0}^{\infty}\left( \frac{\mp j\Omega}{\ii\omega_n \mp E} \right)^k.
\eea
Therefore, after the eigenvector Eq.~\eqref{eq_ApJeigen}, we obtain
\bea
\frac{u_{\alpha_{\ell}}^{a}(\mathbf{x}_1)\overline{u}^{b}_{\alpha_{\ell}}(\mathbf{x}_2)}{\ii\omega_n - E + j\Omega} &=& \frac{1}{\ii\omega_n - E}\sum_{k=0}^{\infty}\left(- \frac{j\Omega}{\ii\omega_n - E} \right)^k u_{\alpha_{\ell}}^{a}(\mathbf{x}_1)\overline{u}^{b}_{\alpha_{\ell}}(\mathbf{x}_2)\nn\\
&=& \frac{1}{\ii\omega_n - E}\sum_{k=0}^{\infty}\left( \frac{-\hat{J}_z\Omega}{\ii\omega_n - E} \right)^k u_{\alpha_{\ell}}^{a}(\mathbf{x}_1)\overline{u}^{b}_{\alpha_{\ell}}(\mathbf{x}_2),
\eea
and similarly
\bea
\frac{v_{\alpha_{\ell}}^{a}(\mathbf{x}_1)\overline{v}^{b}_{\alpha_{\ell}}(\mathbf{x}_2)}{\ii\omega_n + E - j\Omega} &=& \frac{1}{\ii\omega_n + E}\sum_{k=0}^{\infty}\left( \frac{j\Omega}{\ii\omega_n + E} \right)^k v_{\alpha_{\ell}}^{a}(\mathbf{x}_1)\overline{v}^{b}_{\alpha_{\ell}}(\mathbf{x}_2)\nn\\
&=& \frac{1}{\ii\omega_n + E}\sum_{k=0}^{\infty}\left( \frac{-\hat{J}_z\Omega}{\ii\omega_n + E} \right)^k v_{\alpha_{\ell}}^{a}(\mathbf{x}_1)\overline{v}^{b}_{\alpha_{\ell}}(\mathbf{x}_2).
\eea
Therefore, substituting the identities above into Eq.~\eqref{eq_Apspectral}, we obtain
\bea
S'_{ab}(\mathbf{x}_1,\mathbf{x}_2,\ii\omega_n) &=& -\int_{m_f}^{\infty}dE \left[\sum_{k=0}^{\infty} \frac{\left(-\Omega \hat{J}_z\right)^k}{\left(\ii\omega_n - E\right)^{k+1}}  A_{ab}(\mathbf{x}_1,\mathbf{x}_2)+ \sum_{k=0}^{\infty} \frac{\left(-\Omega\hat{J}_z\right)^k}{\left(\ii\omega_n - E\right)^{k+1}} B_{ab}(\mathbf{x}_1,\mathbf{x}_2)\right],
\eea
where we defined~\cite{Vilenkin_PhysRevD.21.2260}
\bea
A_{ab}(\mathbf{x}_1,\mathbf{x}_2) &=&
\sum_{\ell}\sum_h\int_{-p}^{p}dp_z u_{\alpha_{\ell}}^{a}(\mathbf{x}_1)\overline{u}^{b}_{\alpha_{\ell}}(\mathbf{x}_2),\nn\\
B_{ab}(\mathbf{x}_1,\mathbf{x}_2) &=&
\sum_{\ell}\sum_h\int_{-p}^{p}dp_z v_{\alpha_{\ell}}^{a}(\mathbf{x}_1)\overline{v}^{b}_{\alpha_{\ell}}(\mathbf{x}_2).
\eea

Let us notice the operator identity
\bea
\sum_{k=0}^{\infty} \frac{\left(-\Omega \hat{J}_z\right)^k}{\left(\ii\omega_n \pm E\right)^{k+1}} &=& \sum_{k=0}^{\infty}\left(  -\Omega \hat{J}_z\right)^k \frac{(-1)^k}{k!}\frac{\partial^k}{\partial(\ii\omega_n)^k}\left( \frac{1}{\ii\omega_n \pm E} \right)\nn\\
&=& \exp\left[ \Omega \hat{J}_z\frac{\partial}{\partial(\ii\omega_n)}\right] \left( \frac{1}{\ii\omega_n \pm E} \right)
\eea

Therefore, we conclude the exact relation~\cite{Vilenkin_PhysRevD.21.2260}
\bea
S'\left( \mathbf{x}_1,\mathbf{x}_2,\ii\omega_n \right) &=&
e^{ \Omega \hat{J}_z\frac{\partial}{\partial(\ii\omega_n)}} \left[ -\int_{m_f}^{\infty}dE \left(  \frac{A(\mathbf{x}_1,\mathbf{x}_2)}{\ii\omega_n - E} +  \frac{B(\mathbf{x}_1,\mathbf{x}_2)}{\ii\omega_n + E}\right)\right] \nn\\
&=&e^{ \Omega \hat{J}_z\frac{\partial}{\partial(\ii\omega_n)}} S'_0\left( \mathbf{x}_1,\mathbf{x}_2,\ii\omega_n \right),
\eea
where $S'_0\left( \mathbf{x}_1,\mathbf{x}_2,\ii\omega_n \right)$ represents the Fermion propagator, in a non-rotating ($\Omega=0$) system, in cylindrical coordinates. On the other hand, we know that the free propagator, in cartesian coordinates, is given by the expression
\bea
S_0(\mathbf{x}_1,\mathbf{x}_2,\ii\omega_n) = -\int\frac{d^3 p}{(2\pi)^3} e^{\ii\mathbf{p}\cdot\left( \mathbf{x}_1 - \mathbf{x}_2 \right)}S_0\left(\mathbf{p},\ii\omega_n\right),
\eea
where we defined
\bea
S_0\left(\mathbf{p},\ii\omega_n\right) = \frac{\gamma^0\left(\ii\omega_n \right) - \vec{\gamma}\cdot\mathbf{p}+ m_f}{\left(  \ii\omega_n\right)^2 - \mathbf{p}^2 - m_f^2}
\eea
as the non-rotating ($\Omega = 0$) free Fermion propagator in Matsubara-momentum space, with $\gamma^{\mu}$ Dirac matrices.

Since the polar coordinates $(\rho,z,\varphi)$ are rotated with respect to the cartesian coordinates $(x,y,z)$ by an angle $\varphi$ around the z-axis, the two definitions of the free propagator are mutually related by a unitary transformation, represented by a spinor rotation around the $z$-axis, i.e. $\hat{U}(\varphi) = \exp\left( -\ii\varphi \hat{J}_z  \right)$, such that
\bea
S'_0(\mathbf{x}_1,\mathbf{x}_2,\omega_n) = \hat{U}(\varphi_1)S_0(\mathbf{x}_1,\mathbf{x}_2,\omega_n)\hat{U}^{\dagger}(\varphi_2).
\label{eq_S0rot}
\eea
 The same relation holds for the rotating propagators in cartesian $S(\mathbf{x}_1,\mathbf{x}_2,\omega_n)$ and cylindrical $S'(\mathbf{x}_1,\mathbf{x}_2,\omega_n)$ coordinates, respectively, such that by expressing the inverse relation above, we have
\bea
S(\mathbf{x}_1,\mathbf{x}_2,\omega_n) &=& \hat{U}^{\dagger}(\varphi_1)S'(\mathbf{x}_1,\mathbf{x}_2,\omega_n)\hat{U}(\varphi_2)\nn\\
&=& \hat{U}^{\dagger}(\varphi_1)e^{\Omega\hat{J}_z\frac{\partial}{\partial(\ii\omega_n)}}  S'_0(\mathbf{x}_1,\mathbf{x}_2,\omega_n) \hat{U}(\varphi_2)\nn\\
&=& e^{\Omega\hat{J}_z\frac{\partial}{\partial(\ii\omega_n)}} 
\hat{U}^{\dagger}(\varphi_1)S'_0(\mathbf{x}_1,\mathbf{x}_2,\omega_n) \hat{U}(\varphi_2)\nn\\
&=& e^{\Omega\hat{J}_z\frac{\partial}{\partial(\ii\omega_n)}} S_0(\mathbf{x}_1,\mathbf{x}_2,\omega_n).
\label{eq_ASSp}
\eea
Here, in the last step we used the fact that $\left[ \hat{U}^{\dagger}(\varphi_1),\hat{J}_z \right] = 0$ to commute the exponential operator, and the unitary property $\hat{U}^{\dagger}(\varphi)\hat{U}(\varphi) = \hat{U}(\varphi)\hat{U}^{\dagger}(\varphi) = 1$ to solve for $S_0\left( \mathbf{x}_1,\mathbf{x}_2,\ii\omega_n \right)$ from Eq.~\eqref{eq_S0rot}. This result can be cast in a more general form, after noticing that $\Omega \hat{J}_z = \mathbf{\Omega}\cdot\hat{\mathbf{J}} = \mathbf{\Omega}\cdot\left( \hat{\mathbf{L}} + \frac{1}{2}\hat{\mathbf{\Sigma}} \right)$, thus recovering the expression in Eq.~(54) of Vilenkin~\cite{Vilenkin_PhysRevD.21.2260}
\bea
S(\mathbf{x}_1,\mathbf{x}_2,\omega_n) = \exp\left[ \mathbf{\Omega}\cdot\left( \mathbf{L} + \frac{1}{2}\mathbf{\Sigma} \right)\frac{\partial}{\partial(\ii\omega_n)} \right]S_0(\mathbf{x}_1,\mathbf{x}_2,\omega_n).
\label{eq_ApSOmega}
\eea
An important remark about Eq.~\eqref{eq_ApSOmega} is that it trivially reduces, as it should, to the free non-rotating case in the limit $\mathbf{\Omega}\rightarrow 0$, a property that is clearly not satisfied by the expression reported in Ref.~\cite{PhysRevD.103.076021}. On the other hand, even though Eq.~\eqref{eq_ApSOmega} is an exact result, it represents only a formal expression and further algebra and approximations are required to apply it in any practical computation. In order to proceed, we shall first use the expansion of the Fourier exponential in terms of Bessel functions, as follows (for $R = |\mathbf{x}_{\perp,1}-\mathbf{x}_{\perp,2}|$)
\bea
e^{\ii \mathbf{p}\cdot\left( \mathbf{x}_1 - \mathbf{x}_2 \right)} = e^{\ii p_z (z_1 - z_2)} e^{\ii p_{\perp} R \cos(\varphi_1 - \varphi_2)} = e^{\ii p_z (z_1 - z_2)}\sum_{\ell = -\infty}^{\infty}\ii^{\ell}e^{\ii\ell\left( \varphi_1 - \varphi_2 \right)} J_{\ell}\left( p_{\perp}R \right).
\label{eq_ApBessel}
\eea
Specializing ourselves, as shown in Fig.~\ref{fig:coordSystemOmega}, to a coordinate system with the $z$-direction parallel to the vorticity  $\mathbf{\Omega} = \hat{z}\Omega$,
we have 
\bea
S(\mathbf{x}_1,\mathbf{x}_2,\ii\omega_n) &=& -\int\frac{d^3 p}{(2\pi)^3}e^{\left[ \Omega\left( \hat{L}_z + \frac{1}{2}\hat{\Sigma}_z \right)\frac{\partial}{\partial(\ii\omega_n)} \right]}  e^{\ii\mathbf{p}\cdot\left( \mathbf{x}_1 - \mathbf{x}_2 \right)}S_0\left(\mathbf{p},\ii\omega_n\right)\nn\\
&=& -\int\frac{d^3 p}{(2\pi)^3} e^{\ii p_z (z_1 - z_2)} \sum_{\ell = -\infty}^{\infty}\ii^{\ell}e^{\ii\ell\left( \varphi_1 - \varphi_2 \right)} J_{\ell}\left( p_{\perp}R \right) e^{ \left[\Omega\left( \ell + \frac{1}{2}\hat{\Sigma}_z \right)\frac{\partial}{\partial(\ii\omega_n)}\right]}S_0\left(\mathbf{p},\ii\omega_n\right),
\label{eq_ApSOmega2}
\eea
where we used the fact that $e^{\ii \ell(\varphi_1-\varphi_2)}$ are eigenfunctions of $\hat{L}_z  = -\ii\partial_{\varphi_1}$ with eigenvalue $\ell$.

We now introduce the helicity projectors defined in the main text,
\bea
\hat{\mathcal{O}}^{(\pm)} = \frac{\mathbb{1} \pm \hat{\Sigma}^z}{2},
\eea
in order to perform a spectral decomposition of the exponential operator defined in Eq.~\eqref{eq_ApSOmega2} over the subspaces with spin components $+1/2$ and $-1/2$, respectively, 
\bea
e^{ \left[\Omega\left( \ell + \frac{1}{2}\hat{\Sigma}_z \right)\frac{\partial}{\partial(\ii\omega_n)}\right]}S_0\left(\mathbf{p},\ii\omega_n\right) &=& \left[\mathcal{O}^{(+)}e^{ \left[\Omega\left( \ell + \frac{1}{2} \right)\frac{\partial}{\partial(\ii\omega_n)}\right]} + \mathcal{O}^{(-)}e^{ \left[\Omega\left( \ell - \frac{1}{2} \right)\frac{\partial}{\partial(\ii\omega_n)}\right]}\right]S_0\left(\mathbf{p},\ii\omega_n\right)\nn\\
&=& \mathcal{O}^{(+)}e^{ \left[\Omega\left( \ell + \frac{1}{2} \right)\frac{\partial}{\partial(\ii\omega_n)}\right]}S_0\left(\mathbf{p},\ii\omega_n\right) + \mathcal{O}^{(-)}e^{ \left[\Omega\left( \ell - \frac{1}{2} \right)\frac{\partial}{\partial(\ii\omega_n)}\right]}S_0\left(\mathbf{p},\ii\omega_n\right)\nn\\
&=& \mathcal{O}^{(+)} S_0\left(\mathbf{p},\ii\omega_n + \Omega(\ell + \frac{1}{2})\right) + \mathcal{O}^{(-)} S_0\left(\mathbf{p},\ii\omega_n + \Omega(\ell - \frac{1}{2})\right),
\label{eq_ApDiffProj}
\eea
where in the last step we applied the trivial definition of the differential operator that generates finite translations $\exp(a\partial_x)$,
\bea
e^{ a \partial_x} f(x) = f(x) + a f'(x) + \ldots = \sum_{n=0}^{\infty}\frac{a^n}{n!}f^{(n)}(x) = f(x+a).
\eea
By finally inserting Eq.~\eqref{eq_ApDiffProj} into Eq.~\eqref{eq_ApSOmega2}, we obtain the explicit expression for the Fermion operator under rotation $\Omega \ne 0$,
\bea
S(\mathbf{x}_1,\mathbf{x}_2,\ii\omega_n) 
&=& -\int\frac{d^3 p}{(2\pi)^3} e^{\ii p_z (z_1 - z_2)} \sum_{\ell = -\infty}^{\infty}\ii^{\ell}e^{\ii\ell\left( \varphi_1 - \varphi_2 \right)} J_{\ell}\left( p_{\perp}R \right) \left[  \mathcal{O}^{(+)} S_0\left(\mathbf{p},\ii\omega_n + \Omega(\ell + \frac{1}{2})\right)\right.\nn\\
&&\left.+ \mathcal{O}^{(-)} S_0\left(\mathbf{p},\ii\omega_n + \Omega(\ell - \frac{1}{2})\right)\right]
\eea
This last expression is exact, and clearly provides a decomposition into spin $\pm1/2$ and orbital angular momentum $\ell\in\mathbb{Z}$ channels, with the total angular momentum eigenvalues $j = \ell \pm 1/2$ coupling as a shift in the Matsubara frequencies $\ii\omega_n \rightarrow \ii\omega_n + \Omega j$. We proceed further with an approximation, which is to include in the brackets only the lowest orbital momentum mode $\ell = 0$. Invoking again the identity Eq.~\eqref{eq_ApBessel}, this leads to the approximate form
\bea
S(\mathbf{x}_1,\mathbf{x}_2,\ii\omega_n) &=& -\int\frac{d^3 p}{(2\pi)^3} e^{\ii \mathbf{p}\cdot\left(\mathbf{x}_1 - \mathbf{x}_2 \right)} \left[  \mathcal{O}^{(+)} S_0\left(\mathbf{p},\ii\omega_n + \frac{\Omega}{2}\right)+ \mathcal{O}^{(-)} S_0\left(\mathbf{p},\ii\omega_n - \frac{\Omega}{2}\right)\right]\nn\\ 
&\equiv& -\int\frac{d^3 p}{(2\pi)^3} e^{\ii \mathbf{p}\cdot\left(\mathbf{x}_1 - \mathbf{x}_2 \right)} S(\mathbf{p},\ii\omega_n)
\eea
where we defined the rotating Fermion propagator in momentum-Matsubara space
\bea
S(\mathbf{p},\ii\omega_n) &\equiv& \mathcal{O}^{(+)} S_0\left(\mathbf{p},\ii\omega_n + \frac{\Omega}{2}\right)+ \mathcal{O}^{(-)} S_0\left(\mathbf{p},\ii\omega_n - \frac{\Omega}{2}\right)\nn\\
&=& \mathcal{O}^{(+)} \frac{\slashed{p}_+ + m_f}{p_{+}^2 - m_f^2} + \mathcal{O}^{(-)} \frac{\slashed{p}_- + m_f}{p_{-}^2 - m_f^2},
\eea
with $p_{\sigma} = \left( \ii\omega_n + \sigma\Omega/2,\mathbf{p} \right)$, for $\sigma = \pm$ representing the spin projection components.
}

\section{Computation of $g_{\mu\nu}\Pi^{\mu\nu}$}\label{Ap:ComputationOfPi}
From Eq.~\eqref{eq:PolTR} it is clear that
\bea
\Pi^{\mu\nu}=\frac{\ii q_f^2}{2}\int\frac{d^4k}{(2\pi)^4}\text{Tr}\Big\{\gn S\left(k\right)\gm S(k-p)\Big\}.
\eea

According to the definition, the fermion propagator can be written as
\bea
S(p)
&=&\Op{+}\frac{\slashed{p}_++m_f}{p_+^2-m_f^2+\ii\epsilon}+\Op{-}\frac{\slashed{p}_-+m_f}{p_-^2-m_f^2+\ii\epsilon}
\eea
and, similarly,
\bea
S(k-p)
&=&\Op{+}\frac{\left(\slashed{k}-\slashed{p}\right)_++m_f}{\left(k - p\right)_+^2-m_f^2+\ii\epsilon}+\Op{-}\frac{\left(\slashed{k}-\slashed{p}\right)_-+m_f}{\left(k-p\right)_-^2-m_f^2+\ii\epsilon}
\eea
where we have defined
\bea
p_{\sigma} &=& \left( p^0 + \sigma \frac{\Omega}{2},\mathbf{p}  \right)\nn\\
\left(k - p\right)_{\sigma} &=& \left( k^0 - p^0 + \sigma\frac{\Omega}{2},\mathbf{k} - \mathbf{p} \right).
\eea

{\section{Computation of the Dirac trace}\label{AppD}

In this section, we explicitly compute the trace over Dirac matrices that is involved in the definition of the photon polarization tensor,
\bea
\text{Tr}\left\{\ii q_f\gn \ii S(k)\ii q_f\gm \ii S(q)\right\},
\eea
where $q=k-p$ and it is easy to show that the $\text{C.C.}$ contribution is given by
\bea
\text{Tr}\left\{\ii q_f\gn \ii S(-q)\ii q_f\gm \ii S(-k)\right\}. 
\eea

From the expression of the propagator
\bea
S(p)=\sum_{s=\pm1}\Op{s}\frac{\slashed{p}_s+m_f}{p_s^2-m_f^2+\ii\epsilon},
\eea
we have
\bea
&&g_{\mu\nu}\left[\text{Tr}\left\{\ii q_f\gn \ii S(k)\ii q_f\gm \ii S(q)\right\}\right]\nn\\
&=&8q_f^2\left[\frac{m_f^2-k_+\cdot q_-}{(k_+^2-m_f^2)(q_-^2-m_f^2)}+\frac{m_f^2-k_-\cdot q_+}{(k_-^2-m_f^2)(q_+^2-m_f^2)}+\frac{m_f^2}{(k_+^2-m_f^2)(q_+^2-m_f^2)}+\frac{m_f^2}{(k_-^2-m_f^2)(q_-^2-m_f^2)}\right],
\eea
where $q=k-p$. Therefore, the photon polarizatin tensor fully contracted with the metric tensor reduces to
\bea
g_{\mu\nu}\Pi^{\mu\nu}(\omega)=4 q_f^2\sum_{\sigma=\pm1}\left(\mathcal{I}_\sigma+\mathcal{J}_{\sigma}\right),
\eea
where we defined the scalar coefficients
\begin{subequations}
\bea
\mathcal{I}_\sigma\equiv\ii\int\frac{d^4k}{(2\pi)^4}\frac{m_f^2-k_\sigma\cdot(k-p)_{-\sigma}}{(k_\sigma^2-m_f^2)\left[(k-p)_{-\sigma}^2-m_f^2\right]},
\eea
and
\bea
\mathcal{J}_\sigma\equiv\ii\int\frac{d^4k}{(2\pi)^4}\frac{m_f^2}{(k_\sigma^2-m_f^2)\left[(k-p)_{\sigma}^2-m_f^2\right]}.
\eea
\end{subequations}
}
\subsection{Matsubara sums}\label{Ap:MatsubaraSums}
At finite temperature, the temporal component of the momenta is rotated onto the imaginary axis, followed by a discretization in terms of Matsubara frequencies, according to
\bea
k_0 &\rightarrow & \ii \omega_n = \ii (2 n + 1)\pi T\nn\\
p_0 &\rightarrow & \ii \nu_l = \ii 2\pi l T,
\eea
where it is taken into account that $k$ corresponds to the fermionic propagator (odd Matsubara frequencies), whereas $p$ denotes the momentum of the external photon (even Matsubara frequencies).

To perform the Matsubara sum over $\omega_n$, one first identifies the poles of the denominators. There are two distinct types of denominators, namely,
\begin{subequations}
\bea
(k^2_\sigma-m_f^2)((k-p)_{-\sigma}^2-m_f^2)\to\left[ \left( \ii\omega_n +\sigma\Omega/2 \right)^2 - E_{k}^2  \right] \left[ \left( \ii\omega_n - \ii\nu_l-\sigma\Omega/2 \right)^2 - E_{kp}^2 \right].
\label{eq:denominadorcomun2}
\eea
and
\bea
(k_\sigma^2-m_f^2)((k-p)_\sigma^2-m_f^2)\to \left[ \left( \ii\omega_n +\sigma\Omega/2 \right)^2 - E_{k}^2  \right] \left[ \left( \ii\omega_n - \ii\nu_l+\sigma\Omega/2 \right)^2 - E_{kp}^2 \right],
\label{eq:denominadorcomun1}
\eea
\end{subequations}

Here, we have defined
\bea
E_{k} &\equiv& \sqrt{\mathbf{k}^2 + m_f^2}\nn\\
E_{kp} &\equiv& \sqrt{\left(\mathbf{k}-\mathbf{p}\right)^2 + m_f^2}.
\eea

We begin by computing $\mathcal{I}_\sigma$:
\bea
\mathcal{I}_\sigma\equiv\ii\int\frac{d^4k}{(2\pi)^4}\frac{m_f^2-k_\sigma\cdot(k-p)_{-\sigma}}{(k^2_\sigma-m_f^2)((k-p)_{-\sigma}^2-m_f^2)}.
\eea
For $k_{\sigma}\cdot (k - p)_{-\sigma} = \left( \ii\omega_n + \sigma\Omega/2 \right)\left(\ii\omega_n -\ii\nu_l -\sigma\Omega/2  \right) - \mathbf{k}\cdot\left( \mathbf{k} - \mathbf{p}\right)$, this expression can be rewritten as
\bea
\mathcal{I}_\sigma &=& \ii (\ii T) \int\frac{d^3 k}{(2\pi)^3} \sum_{n=-\infty}^{+\infty} 
\frac{m_f^2-\left( \ii\omega_n + \sigma\Omega/2 \right)\left(\ii\omega_n -\ii\nu_l -\sigma\Omega/2  \right) +\,\mathbf{k}\cdot\left( \mathbf{k} - \mathbf{p}\right)}{\left[ \left( \ii\omega_n +\sigma\Omega/2 \right)^2 - E_{k}^2  \right] \left[ \left( \ii\omega_n - \ii\nu_l -\sigma\Omega/2 \right)^2 - E_{kp}^2 \right]}\nn\\
&=& \ii \int\frac{d^3 k}{(2\pi)^3} S_{\sigma}^\mathcal{I}(E_k,E_{kp},\Omega).
\eea

The Matsubara sum is evaluated using the standard contour integration method, by choosing a meromorphic function with infinitely many poles at the discrete odd Matsubara frequencies. The Fermi-Dirac distribution fulfills this property. Therefore, we define
\bea
S_{\sigma}^\mathcal{I}(E_k,E_{kp},\Omega) &=& \ii T \sum_{n=-\infty}^{+\infty} \frac{m_f^2-\left( \ii\omega_n + \sigma\Omega/2 \right)\left(\ii\omega_n -\ii\nu_l -\sigma\Omega/2  \right) +\,\mathbf{k}\cdot\left( \mathbf{k} - \mathbf{p}\right)}{\left[ \left( \ii\omega_n +\sigma\Omega/2 \right)^2 - E_{k}^2  \right] \left[ \left( \ii\omega_n - \ii\nu_l -\sigma\Omega/2 \right)^2 - E_{kp}^2 \right]}\nn\\
&=&-\ii\oint_{C}\frac{dz}{2\pi\ii}\frac{1}{e^{\beta z} + 1}\frac{m_f^2-\left( z + \sigma\Omega/2 \right)\left(z -\ii\nu_l -\sigma\Omega/2  \right) +\mathbf{k}\cdot\left( \mathbf{k} - \mathbf{p}\right)}{\left[ \left( z +\sigma\Omega/2 \right)^2 - E_{k}^2  \right] \left[ \left( z - \ii\nu_l -\sigma\Omega/2 \right)^2 - E_{kp}^2 \right]},
\eea
where $C$ is a contour excluding the imaginary axis, which contains the sequence of simple poles of the Fermi-Dirac distribution $n_\text{F}(\beta z) = (e^{\beta z}+1)^{-1}$ at $z_n = \ii\omega_n$, and provided by the fact that
\bea
\text{Res}\left[\frac{-\beta}{e^{\beta z}+1}\right]_{z=\ii\omega_n}=1.
\eea

The denominator has four simple poles located at
\bea
z_1 &=& -\sigma \frac{\Omega}{2} + E_k\nn\\
z_2 &=& -\sigma \frac{\Omega}{2} - E_k\nn\\
z_3 &=& \sigma \frac{\Omega}{2} - E_{kp}+\ii \nu_l\nn\\
z_4 &=& \sigma \frac{\Omega}{2} + E_{kp}+\ii \nu_l.
\eea

By Cauchy's theorem, the Matsubara sum can be written as
\bea
S_{\sigma}^\mathcal{I}(E_k,E_{kp},\Omega) &=& -\ii\sum_{j = 1}^{4}n_\text{F}(\beta z_j )\text{Res}\left(\frac{m_f^2-\left( z + \sigma\Omega/2 \right)\left(z -\ii\nu_l -\sigma\Omega/2  \right) +\mathbf{k}\cdot\left( \mathbf{k} - \mathbf{p}\right)}{\left[ \left( z +\sigma\Omega/2 \right)^2 - E_{k}^2  \right] \left[ \left( z - \ii\nu_l -\sigma\Omega/2 \right)^2 - E_{kp}^2 \right]} \right)_{z=z_j}\\
&=&-\ii\sum_{s=\pm1}\frac{1}{2E_k}s\,n_\text{F}\left[\beta\left(sE_k-\frac{\sigma\Omega}{2}\right)\right]\frac{m_f^2-sE_k(sE_k-\sigma\Omega-\ii\nu_l)+\mathbf{k}\cdot\left( \mathbf{k} - \mathbf{p}\right)}{\left[E_k+E_{kp}-s\left(\sigma\Omega+\ii\nu_l\right)\right]\left[E_k-E_{kp}-s\left(\sigma\Omega+\ii\nu_l\right)\right]}\nn\\
&-&\ii\sum_{s=\pm1}\frac{1}{2E_{kp}}s\,n_\text{F}\left[\beta\left(sE_{kp}+\frac{\sigma\Omega}{2}+\ii\nu_l\right)\right]\frac{m_f^2-sE_{kp}(sE_{kp}+\sigma\Omega+\ii\nu_l)+\mathbf{k}\cdot\left( \mathbf{k} - \mathbf{p}\right)}{\left[E_{kp}+E_k+s\left(\sigma\Omega+\ii\nu_l\right)\right]\left[E_{kp}-E_k+s\left(\sigma\Omega+\ii\nu_l\right)\right]}.\nn
\eea

Here, $n_\text{F}(x)\equiv (1+e^x)^{-1}$ is the Fermi-Dirac distribution. For bosonic Matsubara frequencies $\ii\nu_l = \ii 2\pi l T$, the Fermi distribution satisfies the trivial property
\bea
n_\text{F}(\beta(x + \ii\nu_l)) = n_\text{F}(\beta x),
\label{eq:NFtrivial}
\eea
which allows the Matsubara sum to be written in the simplified form
\bea
S_{\sigma}^\mathcal{I}(E_k,E_{kp},\Omega) &=& -\ii\sum_{s=\pm1}\frac{1}{2E_k}s\,n_\text{F}\left[\beta\left(sE_k-\frac{\sigma\Omega}{2}\right)\right]\frac{m_f^2-sE_k(sE_k-\sigma\Omega-\ii\nu_l)+\mathbf{k}\cdot\left( \mathbf{k} - \mathbf{p}\right)}{\left[E_k+E_{kp}-s\left(\sigma\Omega+\ii\nu_l\right)\right]\left[E_k-E_{kp}-s\left(\sigma\Omega+\ii\nu_l\right)\right]}\nn\\
&-&\ii\sum_{s=\pm1}\frac{1}{2E_{kp}}s\,n_\text{F}\left[\beta\left(sE_{kp}+\frac{\sigma\Omega}{2}\right)\right]\frac{m_f^2-sE_{kp}(sE_{kp}+\sigma\Omega+\ii\nu_l)+\mathbf{k}\cdot\left( \mathbf{k} - \mathbf{p}\right)}{\left[E_{kp}+E_k+s\left(\sigma\Omega+\ii\nu_l\right)\right]\left[E_{kp}-E_k+s\left(\sigma\Omega+\ii\nu_l\right)\right]}.
\label{eq:sumK}
\eea

Returning back to the Minkowsky space via analytic continuation $\ii\nu_l\to\wph+\ii\epsilon$ (with $\wph$ the photon's energy), we obtain
\bea
\mathcal{I}_\sigma &=&\frac{1}{2}\int\frac{d^3k}{(2\pi)^3}\sum_{s=\pm1}s\,n_\text{F}\left[\beta\left(sE_k-\frac{\sigma\Omega}{2}\right)\right]\frac{m_f^2-sE_k(sE_k-\sigma\Omega-\wph)+\mathbf{k}\cdot\left( \mathbf{k} - \mathbf{p}\right)}{E_k\left[E_k+E_{kp}-s\left(\sigma\Omega+\wph+\ii\epsilon\right)\right]\left[E_k-E_{kp}-s\left(\sigma\Omega+\wph+\ii\epsilon\right)\right]}\nn\\
&+&\frac{1}{2}\int\frac{d^3k}{(2\pi)^3}\sum_{s=\pm1}s\,n_\text{F}\left[\beta\left(sE_{kp}+\frac{\sigma\Omega}{2}\right)\right]\frac{m_f^2-sE_{kp}(sE_{kp}+\sigma\Omega+\wph)+\mathbf{k}\cdot\left( \mathbf{k} - \mathbf{p}\right)}{E_{kp}\left[E_{kp}+E_k+s\left(\sigma\Omega+\wph+\ii\epsilon\right)\right]\left[E_{kp}-E_k+s\left(\sigma\Omega+\wph+\ii\epsilon\right)\right]}.\nn\\
\eea

Furthermore, for on-shell photons with $\mathbf{p}^2 = \wph^2$, we have
\bea
\mathbf{k}\cdot\left( \mathbf{p} - \mathbf{k} \right) = \frac{1}{2}\left(\wph^2 - E_k^2 - E_{kp}^2 + 2 m_f^2\right),
\eea
which allows us to write
\bea
\mathcal{I}_\sigma &=& \frac{1}{2} \int \frac{d^3 k}{(2\pi)^3} \sum_{s=\pm1} s\, n_\text{F}\left[\beta\left(sE_k-\frac{\sigma\Omega}{2}\right)\right] \frac{m_f^2- s E_k (sE_k - \sigma\Omega - \wph) - \frac{1}{2} \left( \wph^2 - E_k^2 - E_{kp}^2 + 2 m_f^2 \right) }{E_k \left[ E_k + E_{kp} - s(\sigma\Omega + \wph + \ii \epsilon) \right] \left[ E_k - E_{kp} - s(\sigma\Omega + \wph + \ii \epsilon) \right]} \nn\\
&+& \frac{1}{2} \int \frac{d^3 k}{(2\pi)^3} \sum_{s=\pm1} s\, n_\text{F}\left[\beta\left(sE_{kp}+\frac{\sigma\Omega}{2}\right)\right] \frac{m_f^2- s E_{kp} (sE_{kp} + \sigma\Omega + \wph) - \frac{1}{2} \left( \wph^2 - E_k^2 - E_{kp}^2 + 2 m_f^2 \right)}{E_{kp} \left[ E_{kp} + E_k + s(\sigma\Omega + \wph + \ii \epsilon) \right] \left[ E_{kp} - E_k + s(\sigma\Omega + \wph + \ii \epsilon) \right]}.\nn\\
\eea
{
For later convenience, the numerators can be simplified and factored, to obtain the following form
\bea
&&\mathcal{I}_\sigma = \frac{1}{2} \int \frac{d^3 k}{(2\pi)^3} \sum_{s=\pm1} s\, n_\text{F}\left[\beta\left(sE_k-\frac{\sigma\Omega}{2}\right)\right] \frac{- \frac{1}{2}\left[ \left( s E_k - \left( \omega + \sigma\Omega \right) \right)^2  + \omega^2 - \left( \omega + \sigma\Omega \right)^2 - E_{kp}^2 \right] }{E_k \left[ E_k + E_{kp} - s(\sigma\Omega + \wph + \ii \epsilon) \right] \left[ E_k - E_{kp} - s(\sigma\Omega + \wph + \ii \epsilon) \right]} \nn\\
&&+ \frac{1}{2} \int \frac{d^3 k}{(2\pi)^3} \sum_{s=\pm1} s\, n_\text{F}\left[\beta\left(sE_{kp}+\frac{\sigma\Omega}{2}\right)\right] \frac{- \frac{1}{2}\left[(sE_{kp} + \wph + \sigma\Omega)^2 + \wph^2 - (\wph + \sigma\Omega)^2  -  E_k^2 \right]}{E_{kp} \left[ E_{kp} + E_k + s(\sigma\Omega + \wph + \ii \epsilon) \right] \left[ E_{kp} - E_k + s(\sigma\Omega + \wph + \ii \epsilon) \right]}
\eea
}
We now turn to the computation of $\mathcal{J}_\sigma$, defined as
\bea
\mathcal{J}_\sigma \equiv \ii \int \frac{d^4 k}{(2\pi)^4} \frac{m_f^2}{(k_\sigma^2 - m_f^2)((k-p)_\sigma^2 - m_f^2)}.
\eea

By analytic continuation onto Matsubara space as described in the previous case, and using Eq.~\eqref{eq:denominadorcomun1}, this expression becomes
\bea
\mathcal{J}_\sigma &=& \ii (\ii T) \int \frac{d^3 k}{(2\pi)^3} \sum_{n=-\infty}^{+\infty} \frac{m_f^2}{\left[ \left( \ii\omega_n + \sigma\Omega/2 \right)^2 - E_k^2 \right] \left[ \left( \ii\omega_n - \ii\nu_l + \sigma\Omega/2 \right)^2 - E_{kp}^2 \right]} \nn\\
&=& \ii \int \frac{d^3 k}{(2\pi)^3} S_\sigma^\mathcal{J}(E_k, E_{kp}, \Omega),
\label{eq:Jsigma1}
\eea
where we defined the Matsubara sum
\bea
S_\sigma^\mathcal{J}(E_k, E_{kp}, \Omega) &=& \ii T \sum_{n=-\infty}^{+\infty} \frac{ m_f^2}{\left[ \left( \ii\omega_n + \sigma\Omega/2 \right)^2 - E_k^2 \right] \left[ \left( \ii\omega_n - \ii\nu_l + \sigma\Omega/2 \right)^2 - E_{kp}^2 \right]} \nn\\
&=& -\ii \oint_C \frac{dz}{2\pi \ii} \frac{1}{e^{\beta z} + 1} \frac{m_f^2}{\left[ \left( z + \sigma\Omega/2 \right)^2 - E_k^2 \right] \left[ \left( z - \ii\nu_l + \sigma\Omega/2 \right)^2 - E_{kp}^2 \right]}.
\eea

Using the same complex contour integration procedure as in the case of $\mathcal{I}_\sigma$, we identify the poles of the integrand at
\bea
z_1 &=& -\sigma \frac{\Omega}{2} + E_k\nn\\
z_2 &=& -\sigma \frac{\Omega}{2} - E_k\nn\\
z_3 &=& -\sigma \frac{\Omega}{2} - E_{kp} + \ii \nu_l\nn\\
z_4 &=& -\sigma \frac{\Omega}{2} + E_{kp} + \ii \nu_l.
\eea

The Matsubara sum can then be expressed in terms of the residues at these poles:
\bea
S_{\sigma}^\mathcal{J}(E_k,E_{kp},\Omega) &=&- \ii \sum_{j=1}^4 n_\text{F}(\beta z_j) \text{Res} \left( \frac{m_f^2}{\left[ \left( z + \sigma\Omega/2 \right)^2 - E_k^2 \right] \left[ \left( z - \ii\nu_l + \sigma\Omega/2 \right)^2 - E_{kp}^2 \right]} \right)_{z=z_j} \nn\\
&=& -\ii \sum_{s=\pm1} \frac{1}{2E_k} s\, n_\text{F} \left[ \beta \left( s E_k - \frac{\sigma\Omega}{2} \right) \right] \frac{m_f^2}{(E_k + E_{kp} - s \ii \nu_l)(E_k - E_{kp} - s \ii \nu_l)} \nn\\
&-& \ii \sum_{s=\pm1} \frac{1}{2E_{kp}} s\, n_\text{F} \left[ \beta \left( s E_{kp} - \frac{\sigma\Omega}{2} + \ii \nu_l \right) \right] \frac{ m_f^2}{(E_{kp} + E_k + s \ii \nu_l)(E_{kp} - E_k + s \ii \nu_l)}.
\eea

Applying the identity in Eq.~\eqref{eq:NFtrivial} and performing the analytic continuation $\ii \nu_l \rightarrow \wph + \ii \epsilon$ to real frequency space, we obtain the retarded component:
\bea
S_{\sigma}^\mathcal{J}(E_k,E_{kp},\Omega) &=&- \ii \sum_{s=\pm1} \frac{1}{2E_k} s\, n_\text{F} \left[ \beta \left( s E_k - \frac{\sigma\Omega}{2} \right) \right] \frac{ m_f^2}{(E_k + E_{kp} - s (\wph + \ii \epsilon)) (E_k - E_{kp} - s (\wph + \ii \epsilon))} \nn\\
&-& \ii \sum_{s=\pm1} \frac{1}{2E_{kp}} s\, n_\text{F} \left[ \beta \left( s E_{kp} - \frac{\sigma\Omega}{2} \right) \right] \frac{m_f^2}{(E_{kp} + E_k + s (\wph + \ii \epsilon)) (E_{kp} - E_k + s (\wph + \ii \epsilon))}.
\eea

Substituting this result into Eq.~\eqref{eq:Jsigma1} leads to the final expression for $\mathcal{J}_\sigma$:
\bea
\mathcal{J}_\sigma &=& \frac{1}{2} \int \frac{d^3 k}{(2\pi)^3} \sum_{s=\pm1} s\, n_\text{F} \left[ \beta \left( s E_k - \frac{\sigma\Omega}{2} \right) \right] \frac{m_f^2}{E_k (E_k + E_{kp} - s (\wph + \ii \epsilon)) (E_k - E_{kp} - s (\wph + \ii \epsilon))} \nn\\
&+& \frac{1}{2} \int \frac{d^3 k}{(2\pi)^3} \sum_{s=\pm1} s\, n_\text{F} \left[ \beta \left( s E_{kp} - \frac{\sigma\Omega}{2} \right) \right] \frac{m_f^2}{E_{kp} (E_{kp} + E_k + s (\wph + \ii \epsilon)) (E_{kp} - E_k + s (\wph + \ii \epsilon))}.
\eea

\subsection{Momentum integrals}\label{Ap.momentumIntegrals}

Let us first consider the integral
\bea
\mathcal{I}_\sigma &=& \frac{1}{2} \int \frac{d^3 k}{(2\pi)^3} \sum_{s=\pm1} s\, n_\text{F}\left[\beta\left(sE_k-\frac{\sigma\Omega}{2}\right)\right] \frac{- \frac{1}{2}\left[ \left( s E_k - \left( \omega + \sigma\Omega \right) \right)^2  + \omega^2 - \left( \omega + \sigma\Omega \right)^2 - E_{kp}^2 \right] }{E_k \left[ E_k + E_{kp} - s(\sigma\Omega + \wph + \ii \epsilon) \right] \left[ E_k - E_{kp} - s(\sigma\Omega + \wph + \ii \epsilon) \right]} \nn\\
&+& \frac{1}{2} \int \frac{d^3 k}{(2\pi)^3} \sum_{s=\pm1} s\, n_\text{F}\left[\beta\left(sE_{kp}+\frac{\sigma\Omega}{2}\right)\right] \frac{- \frac{1}{2}\left[(sE_{kp} + \wph + \sigma\Omega)^2 + \wph^2 - (\wph + \sigma\Omega)^2  -  E_k^2 \right]}{E_{kp} \left[ E_{kp} + E_k + s(\sigma\Omega + \wph + \ii \epsilon) \right] \left[ E_{kp} - E_k + s(\sigma\Omega + \wph + \ii \epsilon) \right]}
\eea

Moreover, note that for all the denominators:
\bea
&&\frac{1}{\left[E_k-E_{kp}-s\left(\sigma\Omega+\wph+\ii\epsilon\right)\right]\left[E_k+E_{kp}-s\left(\sigma\Omega+\wph+\ii\epsilon\right)\right]}\nn\\
&=&\frac{1}{\left[\wph+\sigma\Omega+s\left(E_{kp}-E_{k}\right)+\ii\epsilon\right]\left[\wph+\sigma\Omega-s\left(E_{k}+E_{kp}\right)+\ii\epsilon\right]},
\eea
and similarly 
\bea
&&\frac{1}{\left[E_{kp}-E_{k}+s\left(\sigma\Omega+\wph+\ii\epsilon\right)\right]\left[E_{kp}+E_{k}+s\left(\sigma\Omega+\wph+\ii\epsilon\right)\right]}\nn\\
&=&\frac{1}{\left[\wph+\sigma\Omega+s(E_{kp}-E_k)+\ii\epsilon\right]\left[\wph+\sigma\Omega+s(E_{kp}+E_k)+\ii\epsilon\right]},
\eea
so that by applying the Plemelj's identity
\bea
\lim_{\epsilon\to0}\frac{1}{(A+\ii\epsilon)(B+\ii\epsilon)}=\text{P.V.}\left(\frac{1}{AB}\right)-\ii\pi\frac{\delta(A)}{B-A}+\ii\pi\frac{\delta(B)}{B-A},
\eea
we get for the imaginary part (which is the only piece necessary to compute the dilepton emission rate):
\bea
\text{Im}\left[\mathcal{I}_\sigma\right]&=&-\frac{\pi}{8}\sum_{s=\pm1}\int\frac{d^3 k}{(2\pi)^3}\frac{ n_\text{F}\left[\beta(sE_k-\sigma\frac{\Omega}{2})\right]}{{ E_{kp}E_{k}}}\Big( \left( s E_k - \left( \omega + \sigma\Omega \right) \right)^2  + \omega^2 - \left( \omega + \sigma\Omega \right)^2 - E_{kp}^2\Big)\nn\\
&&\times\Big(\delta\left[\wph+\sigma\Omega+s\left(E_{kp}-E_{k}\right)\right]-\delta\left[\wph+\sigma\Omega-s\left(E_{k}+E_{kp}\right)\right]\Big)\nn\\
&+&\frac{\pi}{8}\sum_{s=\pm1}\int\frac{d^3 k}{(2\pi)^3}\frac{  n_\text{F}\left[\beta(sE_{kp} + \sigma\frac{\Omega}{2})\right]}{{ E_k E_{kp}}}\Big( (sE_{kp} + \wph + \sigma\Omega)^2 + \wph^2 - (\wph + \sigma\Omega)^2  -  E_k^2 
\Big)\nn\\
&&\times\Big(\delta\left[\wph+\sigma\Omega+s(E_{kp}-E_k)\right]-\delta\left[\wph+\sigma\Omega+s(E_{kp}+E_k)\right]\Big).
\eea
We remark that, given the support imposed by the two delta functions in the first term: $\omega + \sigma\Omega - s E_k = \pm s E_{kp}$, the corresponding numerator reduces to
\bea
\left( s E_k - \left( \omega + \sigma\Omega \right) \right)^2 - E_{kp}^2  + \omega^2 - \left( \omega + \sigma\Omega \right)^2 =  
\omega^2 - \left( \omega + \sigma\Omega \right)^2.
\eea
Similarly, the condition imposed by the support of the second pair of delta functions in the second term: $\omega + \sigma\Omega + s E_{kp} = \pm s E_k$, implies that the numerator in this case reduces to
\bea
(sE_{kp} + \wph + \sigma\Omega)^2 -  E_k^2 + \wph^2 - (\wph + \sigma\Omega)^2   =  
\omega^2 - \left( \omega + \sigma\Omega \right)^2.
\eea
Therefore, extracting this common and momentum-independent factor outside the integrals, we obtain the much simpler expression
\bea
\text{Im}\left[\mathcal{I}_\sigma\right]&=&-\frac{\pi}{8}\left[ \omega^2 - \left(\wph + \sigma\Omega\right)^2\right]\sum_{s=\pm1}\int\frac{d^3 k}{(2\pi)^3}\frac{ n_\text{F}\left[\beta(sE_k-\sigma\frac{\Omega}{2})\right]}{{ E_{kp}E_{k}}}\nn\\
&&\times\Big(\delta\left[\wph+\sigma\Omega+s\left(E_{kp}-E_{k}\right)\right]-\delta\left[\wph+\sigma\Omega-s\left(E_{k}+E_{kp}\right)\right]\Big)\nn\\
&+&\frac{\pi}{8}\left[ \omega^2 - \left(\wph + \sigma\Omega\right)^2\right]\sum_{s=\pm1}\int\frac{d^3 k}{(2\pi)^3}\frac{  n_\text{F}\left[\beta(sE_{kp} + \sigma\frac{\Omega}{2})\right]}{{ E_k E_{kp}}}\nn\\
&&\times\Big(\delta\left[\wph+\sigma\Omega+s(E_{kp}-E_k)\right]-\delta\left[\wph+\sigma\Omega+s(E_{kp}+E_k)\right]\Big).
\label{eq_ImIsigma}
\eea

Here, we defined
\bea
E_k &=& \sqrt{\mathbf{k}^2 + m_f^2} \ge m_f\nn\\
E_{kp}(\alpha) &=& \sqrt{\left( \mathbf{k} - \mathbf{p} \right)^2 + m_f^2} = \sqrt{\mathbf{k}^2 + m_f^2 + \wph^2 - 2\wph k \cos\alpha}\ge m_f,
\label{Ek_Ekp_m}
\eea
since the photon momentum is on-shell $|\mathbf{p}| = \wph > 0$, and we defined $\mathbf{k}\cdot\mathbf{p} = \wph k \cos\alpha$.
In spherical coordinates, we have
\be
d^3 k = 2\pi \sin\alpha d\alpha\,k^2\,dk = 2\pi dx k^2 dk,
\ee
where we defined the auxiliary variable $x = \cos\alpha$, for $-1\le x \le 1$. Therefore, we have
\bea
&&\text{Im}\left[\mathcal{I}_\sigma\right]=-\frac{\pi}{8(2\pi)^2}\left[ \omega^2 - \left(\wph + \sigma\Omega\right)^2\right]\sum_{s=\pm1}\int_0^{\infty}dk k^2\int_{-1}^1dx\Bigg\{\frac{ n_\text{F}\left[\beta(sE_k-\sigma\frac{\Omega}{2})\right]}{{  E_{kp}E_{k}}}\nn\\
&&\times\Big(\delta\left[\wph+\sigma\Omega+s\left(E_{kp}-E_{k}\right)\right]-\delta\left[\wph+\sigma\Omega-s\left(E_{k}+E_{kp}\right)\right]\Big)\nn\\
&-&\frac{  n_\text{F}\left[\beta(sE_{kp} + \sigma\frac{\Omega}{2})\right]}{{ E_k E_{kp}}}\Big(\delta\left[\wph+\sigma\Omega+s(E_{kp}-E_k)\right]-\delta\left[\wph+\sigma\Omega+s(E_{kp}+E_k)\right]\Big)\Bigg\}\nn\\
&&=-\frac{\pi}{8(2\pi)^2}\left[ \omega^2 - \left(\wph + \sigma\Omega\right)^2\right]\sum_{s=\pm1}\int_0^{\infty}dk \frac{k^2}{ E_k}\int_{-1}^1dx\Bigg\{\frac{ n_\text{F}\left[\beta(sE_k-\sigma\frac{\Omega}{2})\right]-n_\text{F}\left[\beta(sE_{kp} + \sigma\frac{\Omega}{2})\right]}{  E_{kp}}\nn\\
&&\times\delta\left[\wph+\sigma\Omega+s\left(E_{kp}-E_{k}\right)\right]\nn\\
&&+\frac{  n_\text{F}\left[\beta(sE_{kp} + \sigma\frac{\Omega}{2})\right] - n_\text{F}\left[\beta(-s E_k-\sigma\frac{\Omega}{2})\right]}{E_{kp}}
\delta\left[\wph+\sigma\Omega+s\left(E_{k}+E_{kp}\right)\right]\Bigg\},
\eea
where in the second step we changed $s\rightarrow-s$ in one of the factors.

Let us first consider the support of the delta functions, that we can express as
\bea
\delta\left[\wph+\sigma\Omega + s\left(E_{kp}(x)\mp E_{k}\right)\right] = \frac{\delta(x - x_0^{\mp})}{\left| \frac{\partial E_{kp}}{\partial x}\right|},
\eea
where
\bea
\left| \frac{\partial E_{kp}}{\partial x}\right| = \left| \frac{-2\wph k}{2\sqrt{\mathbf{k}^2 + m_f^2 + \wph^2 - 2\wph k x}} \right| = \frac{\wph k}{E_{kp}(x)},
\label{eq:delta1supp}
\eea
and $x_0^{\mp}$ are the roots of the equations
\bea
\wph + \sigma\Omega  + s\left( E_{kp}(x_0^{\mp}) \mp E_k  \right)=0
\eea
Solving for $E_{kp}(x)$, we have
\bea
E_{kp}(x_0^{\mp}) = \pm E_k -s\left( \wph + \sigma\Omega \right) \ge {m_f}
\label{eq:Ekp1}
\eea
so that
\bea
\text{Im}\left[\mathcal{I}_\sigma\right]&=&-\frac{\pi}{8(2\pi)^2}\left[ \omega^2 - \left(\wph + \sigma\Omega\right)^2\right]\sum_{s=\pm1}\int_{0}^{\infty}dk \frac{k^2}{ E_k}  \int_{-1}^1dx\nn\\
&&\times\Big(\frac{n_\text{F}\left[\beta(s E_k-\sigma\frac{\Omega}{2})\right] - n_\text{F}\left[\beta \left(s E_{kp}(x) + \sigma\frac{\Omega}{2}\right)\right]}{E_{kp}(x)} \frac{E_{kp}(x_0^-)}{\wph k}\delta\left( x - x_0^-\right)\nn\\
&&+\frac{ n_\text{F}\left[\beta(s E_{kp}(x) +\sigma\frac{\Omega}{2})\right] -  n_\text{F}\left[\beta \left( -s E_{k}-\sigma\frac{\Omega}{2}\right)\right]}{ E_{kp}(x)}\frac{E_{kp}(x_0^+)}{\wph k}\delta\left( x - x_0^+\right)\Big)\nn\\
&=& -\frac{\pi}{8(2\pi)^2}\left[ \omega^2 - \left(\wph + \sigma\Omega\right)^2\right] \sum_{s=\pm1}\int_{0}^{\infty}dk \frac{k}{ E_k}\nn\\
&&\times\Big(
\left\{\frac{n_\text{F}\left[\beta(s E_k-\sigma\frac{\Omega}{2})\right] - n_\text{F}\left[\beta(s E_k - \wph - \sigma\frac{\Omega}{2})\right]}{\wph}\right\}\Theta\left[ E_k - s\left(  \wph + \sigma\Omega\right)-{m_f}\right]\nn\\
&&- \left\{  \frac{n_\text{F}\left[\beta(-s E_k-\sigma\frac{\Omega}{2})\right] -  n_\text{F}\left[\beta(- s E_k - \wph - \sigma\frac{\Omega}{2})\right]}{\wph} \right\}\Theta\left[ -E_k - s\left(  \wph + \sigma\Omega\right)-{m_f}\right]
\Big)
\eea

Finally, since the integrand depends only on the energy $E_k = \sqrt{\mathbf{k}^2 + m^2}$, it is convenient to change the variables for the last integral as follows
\bea
\mathbf{k}^2 = E^2 - m_f^2 \Longrightarrow \frac{k dk}{ E} = dE,\,\,\,\,m_f\le E < \infty,
\label{eq:dkdE}
\eea
and making the change $s\rightarrow-s$ in the second term, we obtain
\bea
\text{Im}\left[\mathcal{I}_\sigma\right]&=& -\frac{\pi}{8(2\pi)^2}\left[ \omega^2 - \left(\wph + \sigma\Omega\right)^2\right] \sum_{s=\pm1}\int_{m_f}^{\infty}dE\Bigg(
\left\{ \frac{n_\text{F}\left[\beta(s E-\sigma\frac{\Omega}{2})\right] - n_\text{F}\left[\beta(s E - \wph - \sigma\frac{\Omega}{2})\right]}{\omega}\right\}\nonumber\\
&&\times\left\{\Theta\left[ E - s\left(  \wph + \sigma\Omega\right)-{m_f}\right]-\Theta\left[ -E + s\left(  \wph + \sigma\Omega\right)-{m_f}\right]\right\}
\Bigg)
\eea

We now turn to the computation of $\mathcal{J}_\sigma$, defined as
\bea
\mathcal{J}_\sigma &=& \frac{1}{2} \int \frac{d^3 k}{(2\pi)^3} \sum_{s=\pm1} s\, n_\text{F} \left[ \beta \left( s E_k - \frac{\sigma\Omega}{2} \right) \right] \frac{m_f^2}{E_k (E_k + E_{kp} - s (\wph + \ii \epsilon)) (E_k - E_{kp} - s (\wph + \ii \epsilon))} \nn\\
&+& \frac{1}{2} \int \frac{d^3 k}{(2\pi)^3} \sum_{s=\pm1} s\, n_\text{F} \left[ \beta \left( s E_{kp} - \frac{\sigma\Omega}{2} \right) \right] \frac{m_f^2}{E_{kp} (E_{kp} + E_k + s (\wph + \ii \epsilon)) (E_{kp} - E_k + s (\wph + \ii \epsilon))}.
\eea

After extracting the imaginary part from Plemelj's identity, we end up with
\bea
\text{Im}\left[\mathcal{J}_\sigma\right] &=& -\frac{\pi}{4}m_f^2\int\frac{d^3k}{(2\pi)^3}\sum_{s=\pm 1} \frac{\,n_\text{F}\left[\beta\left(sE_k-\frac{\sigma\Omega}{2}\right)\right]}{ E_{kp} E_k}\left( \delta\left[ \wph - s\left(  E_k - E_{kp}\right) \right]
-\delta\left[\wph -s\left( E_k + E_{kp}  \right) \right]\right)\nn\\
&+&\frac{\pi}{4}m_f^2\int\frac{d^3 k}{(2\pi)^3}\sum_{s=\pm 1} \frac{\,n_\text{F}\left[\beta\left(sE_{kp}-\frac{\sigma\Omega}{2}\right)\right]}{ E_k E_{kp}}\left( \delta\left[ \wph - s\left(  E_k - E_{kp}\right) \right] -
\delta\left[ \wph + s\left( E_{kp} + E_k \right)\right]\right).
\label{eq_ImJsigma}
\eea

As in the previous case of~~$\mathcal{I}_{\sigma}$, we shall perform the integral in spherical coordinates, and we shall apply the change of variables defined for the delta functions in Eq.~\eqref{eq:delta1supp}. After some simplifications, the expression reduces to
\bea
\text{Im}\left[\mathcal{J}_\sigma\right] &=& -\frac{\pi m_f^2}{4(2\pi)^2}\sum_{s=\pm 1} \int_0^{\infty}dk k^2\int_{-1}^1 dx \frac{\,n_\text{F}\left[\beta\left(sE_k-\frac{\sigma\Omega}{2}\right)\right]}{  E_{kp}(x) E_k}\left( \delta\left[ \wph - s\left(  E_k - E_{kp}(x)\right) \right]
-\delta\left[\wph -s\left( E_k + E_{kp}(x)  \right) \right]\right)\nn\\
&+&\frac{\pi m_f^2}{4(2\pi)^2}\sum_{s=\pm 1} \int_0^{\infty}dk k^2\int_{-1}^1 dx \frac{\,n_\text{F}\left[\beta\left(sE_{kp}(x)-\frac{\sigma\Omega}{2}\right)\right]}{  E_k E_{kp}(x)}
\left( \delta\left[ \wph - s\left(  E_k - E_{kp}\right) \right] -
\delta\left[ \wph + s\left( E_{kp}(x) + E_k \right)\right]\right)\nn\\
\eea

Distributing the factors, and introducing the change $s\rightarrow -s$ in one of the terms, the expression further simplifies to

\bea
\text{Im}\left[\mathcal{J}_\sigma\right] &=& -\frac{ m_f^2\pi}{4(2\pi)^2}\sum_{s=\pm 1} \int_0^{\infty}dk \frac{k^2}{ E_k}\int_{-1}^1 dx \Bigg\{ \frac{\,n_\text{F}\left[\beta\left(sE_k-\frac{\sigma\Omega}{2}\right)\right] - \,n_\text{F}\left[\beta\left(sE_{kp}(x)-\frac{\sigma\Omega}{2}\right)\right]}{E_{kp}(x)}\delta\left[ \wph + s\left(  E_{kp}(x) - E_{k}\right) \right]\nn\\
&&+\frac{\,n_\text{F}\left[\beta\left(s E_{kp}(x)-\frac{\sigma\Omega}{2}\right)\right] - \,n_\text{F}\left[\beta\left(-s E_{k} -\frac{\sigma\Omega}{2}\right)\right]}{E_{kp}(x)}\delta\left[\wph + s\left( E_k + E_{kp}(x)  \right) \right]\Bigg\},
\eea

We analyze the condition imposed by the deltas,
\bea
\delta\left[\wph+s\left( E_{kp}(x) \mp E_k\right)\right] = \frac{\delta(x - x_1^{\mp})}{\left| \frac{\partial E_{kp}}{\partial x}\right|} = \frac{E_{kp}(x_1^{\mp})}{\wph k}\delta(x - x_1^{\mp}),
\eea
where $x_1^{\pm}$ is defined as the solution to the equation
\bea
\wph+s\left(E_{kp}(x_1^{\mp})\mp E_k\right) = 0,
\eea
that corresponds to
\bea
E_{kp}(x_1^{\mp}) = \pm E_k - s \wph   > m_f.
\eea

Performing the change of variables as before, and the required substitutions, we end up with the expression
{
\bea
\text{Im}\left[\mathcal{J}_\sigma\right] &=& -\frac{ m_f^2\pi}{4(2\pi)^2}\sum_{s=\pm 1} \int_0^{\infty}dk \frac{k^2}{ E_k}\int_{-1}^1 dx \Bigg\{ \frac{\,n_\text{F}\left[\beta\left(sE_k-\frac{\sigma\Omega}{2}\right)\right] - \,n_\text{F}\left[\beta\left(s E_{kp}(x)-\frac{\sigma\Omega}{2}\right)\right]}{E_{kp}(x)}\frac{E_{kp}(x_1^{-})}{\wph k}\delta(x - x_1^{-})\nn\\
&&+\frac{\,n_\text{F}\left[\beta\left(sE_{kp}(x)-\frac{\sigma\Omega}{2}\right)\right] - \,n_\text{F}\left[\beta\left(-s E_k-\frac{\sigma\Omega}{2}\right)\right]}{E_{kp}(x)}\frac{E_{kp}(x_1^{+})}{\wph k}\delta(x - x_1^{+})\Bigg\}
\eea
}
By further substituting the identities
\bea
E_{kp}(x_1^{+}) &=& -E_k - s\wph > m_f \nn\\
E_{kp}(x_1^{-}) &=&  E_k -s\wph > m_f 
\eea
in the integral, we arrive at
\bea
\text{Im}\left[\mathcal{J}_\sigma\right] &=& -\frac{ m_f^2\pi}{4\wph(2\pi)^2}\sum_{s=\pm 1} \int_0^{\infty}dk \frac{k}{ E_k}\Bigg\{ \Big(\,n_\text{F}\left[\beta\left(sE_k-\frac{\sigma\Omega}{2}\right)\right] - \,n_\text{F}\left[\beta\left(sE_k - \wph -\frac{\sigma\Omega}{2}\right)\right]\Big)\Theta\left[ E_k - s\wph - m_f\right]\nn\\
&&+\Big( \,n_\text{F}\left[\beta\left(-sE_k-\frac{\sigma\Omega}{2} - \omega\right)\right] - \,n_\text{F}\left[\beta\left(-sE_k -\frac{\sigma\Omega}{2}\right)\right]  \Big)
\Theta\left[ -E_k - s\wph - m_f \right]\Bigg\}
\eea

{Introducing the change $s\rightarrow -s$ in the second term, and making the change of variables defined in Eq.~\eqref{eq:dkdE}, we arrive at
\bea
\text{Im}\left[\mathcal{J}_\sigma\right] &=& -\frac{ m_f^2\pi}{4(2\pi)^2}\sum_{s=\pm 1} \int_{m_f}^{\infty}dE  \Bigg\{\frac{n_\text{F}\left[\beta\left(sE-\frac{\sigma\Omega}{2}\right)\right] - n_\text{F}\left[\beta\left(sE - \wph -\frac{\sigma\Omega}{2}\right)\right]}{\omega}\Bigg\}\nn\\
&&\times\Bigg\{\Theta\left[ E - s\wph - m_f\right]-
\Theta\left[ -E + s\wph - m_f \right]\Bigg\}
\eea

Collecting both contributions, we arrive at the final result
\bea
\text{Im}\left[g_{\mu\nu}\Pi^{\mu\nu}_R(\omega)\right] &=& 4 q_f^2 \sum_{\sigma = \pm 1}\left( \text{Im}\left[\mathcal{I}_{\sigma}\right] + \text{Im}\left[\mathcal{J}_{\sigma}\right]\right)\nn\\
&=& \frac{q_f^2}{8\pi}\left[ 2 m_f^2 \mathcal{F}\left( \beta,\omega,\Omega \right) + \left[(\omega + \sigma\Omega)^2 - \wph^2 \right]\mathcal{G}\left( \beta,\omega,\Omega \right)\right],
\eea
where we defined the functions
\bea
\mathcal{F}\left( \beta,\omega,\Omega \right) = \sum_{\sigma=\pm 1}\sum_{s=\pm 1} \int_{m_f}^{\infty}dE  \,\Delta n_{F}\left(s E,\wph,\sigma\Omega\right)\Bigg\{\Theta\left[ E - s\wph - m_f\right]-
\Theta\left[ -E + s\wph - m_f \right]\Bigg\},
\eea
and
\bea
\mathcal{G}\left( \beta,\omega,\Omega \right) = \sum_{\sigma=\pm1}\sum_{s=\pm1}\int_{m_f}^{\infty}dE
\,\Delta n_{F}\left(s E,\wph,\sigma\Omega\right)\Bigg\{\Theta\left[ E - s\left(  \wph + \sigma\Omega\right)-{m_f}\right]-\Theta\left[ -E + s\left(  \wph + \sigma\Omega\right)-{m_f}\right]
\Bigg\},
\eea
with
\bea
\Delta n_{F}\left(s E,\wph,\sigma\Omega\right) = \frac{n_\text{F}\left[\beta(s E-\sigma\frac{\Omega}{2})\right] - n_\text{F}\left[\beta(s E - \wph - \sigma\frac{\Omega}{2})\right]}{\omega}.
\eea
These integrals can be computed analytically. For this purpose, we first notice that the function in the integrand
\bea
\Theta\left[ E - s\wph - m_f\right]-
\Theta\left[ -E + s\wph - m_f \right] = \left\{ \begin{array}{cc} 
+1, & E > s\omega + m_f\\
0, & s\omega - m_f < E < s\omega + m_f\\-1, & E < s\omega - m_f
\end{array}\right.
\eea
In addition, the lower limit of the imposes the condition $E \ge m_f$. Therefore, if we define the index function
\bea
f_s(E,\omega) = \theta\left[ E - m_f \right]\cdot\left\{ \Theta\left[ E - s\wph - m_f\right]-
\Theta\left[ -E + s\wph - m_f \right] \right\},
\eea
we obtain the following explicit expressions for $s = \pm$, respectively:

For $\omega > 2 m_f$, we have
\bea
f_{+}( E,\omega  ) = \left\{ \begin{array}{cc} +1, & E > m_f + \omega\\
0, & \wph - m_f < E < \wph + m_f\\
-1, & m_f < E < -m_f + \omega\\
0, & E < m_f
\end{array}\right., 
\eea
while for $\omega < 2 m_f$
\bea
f_{+}( E,\omega ) = \left\{ \begin{array}{cc} +1, & E > m_f + \omega\\
0, & E < m_f + \omega
\end{array}\right.. 
\eea
On the other hand, for $s = -1$, we obtain
\bea
f_-(E,\omega) = \left\{  
\begin{array}{cc}
+1, & E > m_f\\
0, & E < m_f
\end{array}
\right..
\eea
Therefore, the integral expression for $\mathcal{F}(\beta,\wph,\Omega)$ can be written as
\bea
\mathcal{F}(\beta,\wph,\Omega) &=& \sum_{\sigma=\pm 1}\sum_{s=\pm 1}\int_{0}^{\infty} dE f_s(E,\wph) \Delta n_{F}\left(s E,\wph,\sigma\Omega\right)\nn\\
&=& \sum_{\sigma = \pm 1}\Bigg\{ 
\Theta(\wph - 2 m_f)\left[\int_{m_f + \wph}^{\infty} dE \Delta n_{F}\left(  E,\wph,\sigma\Omega\right)-\int_{m_f}^{\omega - m_f}dE \Delta n_{F}\left(  E,\wph,\sigma\Omega\right)\right]\nn\\
&&+ \Theta( 2m_f - \omega)\int_{m_f + \wph}^{\infty} dE \Delta n_{F}\left( E,\wph,\sigma\Omega\right)+\int_{m_f }^{\infty} dE \Delta n_{F}\left(- E,\wph,\sigma\Omega\right)
\Bigg\}
\label{eq_intF}
\eea
The integrals to be solved are of the general form (for suitable limits $A$ and $B$)
\begin{eqnarray}
\int_A^B dE\, n_F\left[ \beta \left( s E - b  \right) \right] &=& \int_A^B dE \frac{e^{-\beta\left( s E - b  \right)}}{1 + e^{-\beta\left( s E - b  \right)}} = -\frac{1}{\beta s}\left.\ln\left[ 1 + e^{-\beta\left( s E - b  \right)} \right]\right|_A^B\nn\\
&=& \frac{1}{\beta s}\ln\left[ \frac{1 + e^{-\beta\left( s A - b  \right)} }{1 + e^{-\beta\left( s B - b  \right)}} \right],
\label{eq_intn2}
\end{eqnarray}
and moreover, for the explicit integrand involved in Eq.~\eqref{eq_intF} we have
\bea
\int_A^B dE\, \Delta n_{F}\left(s E,\wph,\sigma\Omega\right) = \frac{1}{\wph \beta s} \ln\left[ \frac{ 1 + e^{-\beta\left( s A - \frac{\sigma\Omega}{2}  \right)}  }{1 + e^{-\beta\left( s A - \omega - \frac{\sigma\Omega}{2}  \right)} } \frac{ 1 + e^{-\beta\left( s B -\omega - \frac{\sigma\Omega}{2}  \right)}  }{1 + e^{-\beta\left( s B - \frac{\sigma\Omega}{2}  \right)} }\right].\nn\\
\eea
Applying this basic formula, along with the identity $1 + e^{-z} = 2 e^{-z/2}\cosh(z/2)$, we obtain
\bea
&&\mathcal{F}(\beta,\wph,\Omega) = \frac{1}{\beta\omega}\sum_{\sigma=\pm 1}\Bigg\{
\Theta(2 m_f - \wph)\ln\left[\frac{e^{-\frac{\beta\omega}{2}}\cosh\left[ \frac{\beta}{2}\left( m_f - \frac{\sigma\Omega}{2} + \wph \right) \right]}
{\cosh\left[ \frac{\beta}{2}\left( m_f - \frac{\sigma\Omega}{2} \right)  \right]} \right]
- \ln\left[\frac{e^{-\frac{\beta\omega}{2}}\cosh\left[ \frac{\beta}{2}\left( m_f + \frac{\sigma\Omega}{2}  \right) \right]}
{\cosh\left[ \frac{\beta}{2}\left( m_f + \frac{\sigma\Omega}{2} + \wph \right)  \right]} \right]\nn\\
&&+\Theta\left( \wph - 2 m_f  \right)
\ln\left[ \frac{e^{-\frac{\beta\wph}{2}}\cosh\left[ \frac{\beta}{2} (m_f - \frac{\sigma\Omega}{2} + \wph )\right] \cosh\left[ \frac{\beta}{2} (m_f - \frac{\sigma\Omega}{2} - \wph )\right] \cosh\left[ \frac{\beta}{2} (m_f + \frac{\sigma\Omega}{2} - \wph )\right]}{\cosh^2\left[ \frac{\beta}{2} (m_f - \frac{\sigma\Omega}{2} )\right] \cosh\left[ \frac{\beta}{2} (m_f + \frac{\sigma\Omega}{2} )\right]  } \right]
\Bigg\} 
\eea
By further applying the identity $\Theta(z) + \Theta(-z) = 1$ to absorb the second term into the first and third one, respectively, we obtain
\bea
&&\mathcal{F}(\beta,\wph,\Omega) = \frac{1}{\beta\omega}\sum_{\sigma=\pm 1}\Bigg\{
\Theta(2 m_f - \wph)\ln\left[\frac{\cosh\left[ \frac{\beta}{2}\left( m_f - \frac{\sigma\Omega}{2} + \wph \right) \right]\cosh\left[ \frac{\beta}{2}\left( m_f + \frac{\sigma\Omega}{2} + \wph \right) \right]}
{\cosh\left[ \frac{\beta}{2}\left( m_f - \frac{\sigma\Omega}{2} \right)  \right]\cosh\left[ \frac{\beta}{2}\left( m_f + \frac{\sigma\Omega}{2} \right)  \right]} \right]\\
&&+\Theta\left( \wph - 2 m_f  \right)
\ln\left[ \frac{\cosh\left[ \frac{\beta}{2} (m_f - \frac{\sigma\Omega}{2} + \wph )\right] \cosh\left[ \frac{\beta}{2} (m_f - \frac{\sigma\Omega}{2} - \wph )\right] \cosh\left[ \frac{\beta}{2} (m_f + \frac{\sigma\Omega}{2} - \wph )\right]\cosh\left[ \frac{\beta}{2} (m_f + \frac{\sigma\Omega}{2} + \wph )\right]}{\cosh^2\left[ \frac{\beta}{2} (m_f - \frac{\sigma\Omega}{2} )\right] \cosh^2\left[ \frac{\beta}{2} (m_f + \frac{\sigma\Omega}{2} )\right]  } \right]
\Bigg\}\nn 
\eea
Clearly, the expression above is {\it{even}} in the index $\sigma = \pm$, and hence the final result becomes
\bea\label{eq_AFfinal}
&&\mathcal{F}(\beta,\wph,\Omega) = 
\Theta(2 m_f - \wph)\frac{2}{\beta\omega}\ln\left[\frac{\cosh\left[ \frac{\beta}{2}\left( m_f - \frac{\Omega}{2} + \wph \right) \right]\cosh\left[ \frac{\beta}{2}\left( m_f + \frac{\Omega}{2} + \wph \right) \right]}
{\cosh\left[ \frac{\beta}{2}\left( m_f - \frac{\Omega}{2} \right)  \right]\cosh\left[ \frac{\beta}{2}\left( m_f + \frac{\Omega}{2} \right)  \right]} \right]\\
&&+\Theta\left( \wph - 2 m_f  \right)
\frac{2}{\beta\omega}\ln\left[ \frac{\cosh\left[ \frac{\beta}{2} (m_f - \frac{\Omega}{2} + \wph )\right] \cosh\left[ \frac{\beta}{2} (m_f - \frac{\Omega}{2} - \wph )\right] \cosh\left[ \frac{\beta}{2} (m_f + \frac{\Omega}{2} - \wph )\right]\cosh\left[ \frac{\beta}{2} (m_f + \frac{\Omega}{2} + \wph )\right]}{\cosh^2\left[ \frac{\beta}{2} (m_f - \frac{\Omega}{2} )\right] \cosh^2\left[ \frac{\beta}{2} (m_f + \frac{\Omega}{2} )\right]  } \right]\nn
\eea

Let us now consider the calculation of $\mathcal{G}\left( \beta,\omega,\Omega \right)$. For this purpose, we remark that the function in the integrand is given in this second case by
\bea
\Theta\left[ E - s\left(\wph + \sigma\Omega\right) - m_f\right]-
\Theta\left[ -E + s\left(\wph + \sigma\Omega\right) - m_f \right] = \left\{ \begin{array}{cc} 
+1, & E > s\left(\wph + \sigma\Omega\right) + m_f\\
0, & s\left(\wph + \sigma\Omega\right) - m_f < E < s\left(\wph + \sigma\Omega\right) + m_f\\
-1, & E < s\left(\wph + \sigma\Omega\right) - m_f
\end{array}\right.
\eea
As in the former case, the lower integration limit imposes the condition $E \ge m_f$. Therefore, we define the index function
\bea
f_s(E,\omega + \sigma\Omega) = \theta\left[ E - m_f \right]\cdot\left\{ \Theta\left[ E - s\left(\wph + \sigma\Omega\right) - m_f\right]-
\Theta\left[ -E + s\left(\wph + \sigma\Omega \right) - m_f \right] \right\}.
\eea
In this second case, we must take into account the possibility that, for $\sigma = -1$, we may reach conditions where $\wph - \Omega < 0$. Therefore, explicit expressions for the case $s = +1$ are as follows:

For $\omega + \sigma\Omega > 2 m_f$,
\bea
f_{+}( E,\omega + \sigma\Omega ) = \left\{ \begin{array}{cc} +1, & E > m_f + \omega + \sigma \Omega\\
0, & -m_f + \wph + \sigma\Omega < E < m_f + \wph + \sigma\Omega\\
-1, & m_f < E < -m_f + \omega + \sigma\Omega\\
0, & E < m_f
\end{array}\right., 
\eea
while for $\wph + \sigma\Omega < 2 m_f$,
\bea
f_{+}( E,\omega + \sigma\Omega ) = \left\{ \begin{array}{cc} +1, & E > m_+^{\sigma}\\
0, & E < m_+^{\sigma}
\end{array}\right.. 
\eea
On the other hand, for the case $s = -1$, we have the following:

For $\omega + \sigma\Omega < - 2  m_f$,
\bea
f_{-}( E,\omega + \sigma\Omega ) = \left\{ \begin{array}{cc} +1, & E > m_f - \left( \wph + \sigma\Omega \right)\\
0, & -m_f -(\wph + \sigma\Omega) < E < m_f - (\wph + \sigma\Omega)\\
-1, & m_f < E < -m_f - \left( \wph + \sigma\Omega \right)\\
0, & E < m_f 
\end{array}\right.. 
\eea
while for $\omega + \sigma\Omega > - 2 m_f$,
\bea
f_{-}( E,\omega + \sigma\Omega ) = \left\{ \begin{array}{cc} +1, & E > m_-^{\sigma}\\
0, & E < m_-^{\sigma} 
\end{array}\right., 
\eea
where we defined (for $s = \pm 1$ and $\sigma = \pm$)
\bea
m_s^{\sigma} = {\text{Max}}\left\{ m_f, m_f + s\left( \wph + \sigma\Omega \right) \right\}.
\eea
These four parameters are completely specified for two particular cases,
\bea
m_+^+ &=& m_f + \wph + \Omega\\
m_{-}^+ &=& m_f.
\eea
Therefore, the integral expression for $\mathcal{G}(\beta,\wph,\Omega)$ can be written as
\bea
\mathcal{G}(\beta,\wph,\Omega) &=& \sum_{\sigma=\pm 1}\sum_{s=\pm 1}\int_{0}^{\infty} dE f_s(E,\wph) \Delta n_{F}\left(s E,\wph,\sigma\Omega\right)\\ 
&=& \sum_{\sigma=\pm 1}\left\{\int_{0}^{\infty} dE f_+(E,\wph) \Delta n_{F}\left( E,\wph,\sigma\Omega\right) + \int_{0}^{\infty} dE f_-(E,\wph) \Delta n_{F}\left( -E,\wph,\sigma\Omega\right)\right\}\nn\\
&=& \sum_{\sigma = \pm 1}\Bigg\{ 
\Theta(\wph + \sigma\Omega - 2 m_f)\left[\int_{m_f + \wph + \sigma\Omega}^{\infty} dE \Delta n_{F}\left(  E,\wph,\sigma\Omega\right)-\int_{m_f}^{\omega + \sigma\Omega - m_f}dE \Delta n_{F}\left(  E,\wph,\sigma\Omega\right)\right]\nn\\
&&+ \Theta( 2m_f - \omega - \sigma\Omega)\int_{m_+^{\sigma}}^{\infty} dE \Delta n_{F}\left( E,\wph,\sigma\Omega\right)+\Theta\left( \wph + \sigma\Omega + 2m_f 
 \right)\int_{m_-^{\sigma} }^{\infty} dE \Delta n_{F}\left(- E,\wph,\sigma\Omega\right)\nn\\
 &&+ \Theta\left( - \wph -\sigma\Omega - 2 m_f \right) \left[ \int_{m_f - (\wph + \sigma\Omega) }^{\infty} dE \Delta n_{F}\left(- E,\wph,\sigma\Omega\right) - \int_{m_f}^{- m_f - (\wph + \sigma\Omega)} dE \Delta n_{F}\left(- E,\wph,\sigma\Omega\right) \right]
\Bigg\}\nn
\label{eq_intG}
\eea
Computing each of the integrals by means of Eq.~\eqref{eq_intn2},
and applying the identity $\Theta(-\omega-\sigma\Omega-2m_f) = 1 - \Theta (2 m_f + \omega + \sigma\Omega)$, we have
\bea
\mathcal{G}(\beta,\wph,\Omega) &=& \frac{1}{\beta\omega}\sum_{\sigma=\pm 1}\Bigg\{
\Theta(\omega + \sigma\Omega - 2 m_f)\ln\left[\frac{e^{-\frac{\beta\omega}{2}}\cosh\left[ \frac{\beta}{2}(m_f + \wph + \frac{\sigma\Omega}{2})\right]\cosh^2\left[\frac{\beta}{2}(m_f - \frac{\sigma\Omega}{2} - \wph) \right] }{\cosh\left[ \frac{\beta}{2}(m_f  + \frac{\sigma\Omega}{2})\right]\cosh^2\left[\frac{\beta}{2}(m_f - \frac{\sigma\Omega}{2} ) \right]} \right]\nn\\
&&+ \Theta(2 m_f - \wph - \sigma\Omega) \ln\left[\frac{e^{-\frac{\beta\omega}{2}}\cosh\left[ \frac{\beta}{2}(m_+^{\sigma} - \frac{\sigma\Omega}{2})\right]}{\cosh\left[ \frac{\beta}{2}(m_+^{\sigma} - \wph  - \frac{\sigma\Omega}{2})\right]} \right]-\Theta(2 m_f + \wph + \sigma\Omega) \ln\left[\frac{e^{-\frac{\beta\omega}{2}}\cosh\left[ \frac{\beta}{2}(m_-^{\sigma} + \frac{\sigma\Omega}{2})\right]}{\cosh\left[ \frac{\beta}{2}(m_-^{\sigma} + \wph  + \frac{\sigma\Omega}{2})\right]} \right]\nn\\
&&-\left[ 1 - \Theta(2m_f + \sigma\Omega + \wph) \right]\ln\left[\frac{e^{-\frac{\beta\omega}{2}}\cosh\left[ \frac{\beta}{2}(m_f - \wph - \frac{\sigma\Omega}{2})\right]\cosh^2\left[\frac{\beta}{2}(m_f + \frac{\sigma\Omega}{2} + \wph) \right] }{\cosh\left[ \frac{\beta}{2}(m_f  - \frac{\sigma\Omega}{2})\right]\cosh^2\left[\frac{\beta}{2}(m_f + \frac{\sigma\Omega}{2} ) \right]} \right]
\Bigg\}
\eea
Finally, making use of the identity $1 = \Theta (\omega+\sigma\Omega - 2m_f) + \Theta(2m_f - \omega - \sigma\Omega)$, it is possible to recombine the terms above into the simplified form
\bea\label{eq_AGfinal}
\mathcal{G}(\beta,\wph,\Omega) &=& \frac{1}{\beta\omega}\sum_{\sigma=\pm 1}\Bigg\{
\Theta(\omega + \sigma\Omega - 2 m_f)\ln\left[\frac{\cosh\left[ \frac{\beta}{2}(m_f + \frac{\sigma\Omega}{2})\right]\cosh\left[\frac{\beta}{2}(m_f - \frac{\sigma\Omega}{2} - \wph) \right] }{\cosh\left[ \frac{\beta}{2}(m_f  - \frac{\sigma\Omega}{2})\right]\cosh\left[\frac{\beta}{2}(m_f + \frac{\sigma\Omega}{2} + \wph ) \right]} \right]\\
&&+ \Theta(2 m_f - \wph - \sigma\Omega) \ln\left[\frac{\cosh\left[ \frac{\beta}{2}(m_+^{\sigma} - \frac{\sigma\Omega}{2})\right]\cosh\left[ \frac{\beta}{2}(m_f - \frac{\sigma\Omega}{2})\right]\cosh^2\left[ \frac{\beta}{2}(m_f + \frac{\sigma\Omega}{2})\right]}{\cosh\left[ \frac{\beta}{2}(m_+^{\sigma} - \wph  - \frac{\sigma\Omega}{2})\right]\cosh\left[ \frac{\beta}{2}(m_f - \frac{\sigma\Omega}{2}-\wph)\right]\cosh^2\left[ \frac{\beta}{2}(m_f + \frac{\sigma\Omega}{2}-\wph)\right]} \right]\nn\\
&&+\Theta(2 m_f + \wph + \sigma\Omega) \ln\left[\frac{\cosh\left[ \frac{\beta}{2}(m_f-\wph - \frac{\sigma\Omega}{2})\right]\cosh^2\left[ \frac{\beta}{2}(m_f + \frac{\sigma\Omega}{2}+\wph)\right]\cosh\left[ \frac{\beta}{2}(m_-^{\sigma} + \frac{\sigma\Omega}{2}+\wph)\right]}{\cosh\left[ \frac{\beta}{2}(m_f -  \frac{\sigma\Omega}{2})\right]\cosh^2\left[ \frac{\beta}{2}(m_f +  \frac{\sigma\Omega}{2})\right]\cosh\left[ \frac{\beta}{2}(m_-^{\sigma} +  \frac{\sigma\Omega}{2})\right]} \right]
\Bigg\}\nn
\eea
\section{The limit $\Omega = 0$}
\label{AppE}
In this section, we shall compare the limit of our analytical expression for the rate, with the result obtained by Weldon~\cite{Weldon_98_PhysRevD.28.2007} for the Fermion propagator coupled to a Yukawa scalar field, that we quote as follows (see Eq.~[2.22] in~\cite{Weldon_98_PhysRevD.28.2007})
\bea
\text{Im}\Pi(\omega,T) &=& \frac{\ii}{2}\text{Disc}\Pi(\omega,T) = \frac{g^2}{2}\int\frac{d^3 k}{(2\pi)^3}\frac{N}{2 E_1 2 E_2}\Bigg[  
\delta(\omega - E_1 - E_2)\left[  (1 - n_1)(1 - n_2) - n_1 n_2\right]\nn\\
&-& \delta(\omega + E_1 - E_2)\left[n_1(1 - n_2) - n_2(1 - n_1) \right] - \delta(\omega-E_1 + E_2)\left[ n_2(1 - n_1) - n_1 (1 - n_2) \right]\nn\\
&+&\delta(\omega + E_1 + E_2)\left[ 
n_1 n_2 - (1 - n_1)(1 - n_2)
\right]
\Bigg],
\label{eq_Weldon}
\eea
with $E_1 \equiv E_{\mathbf{k}} = \sqrt{m_f^2 + \mathbf{k}^2}$, $E_2\equiv E_{kp} = \sqrt{m_f^2 + (\mathbf{k} - \mathbf{p})^2}$, and $n_j\equiv n_F\left( \beta E_j \right)$, respectively.
Since Eq.~\eqref{eq_Weldon} corresponds to the result for Fermions coupled to a Yukawa scalar theory~\cite{Weldon_98_PhysRevD.28.2007}, the overall coefficient will differ from ours due to the Dirac traces involved. Nevertheless, the expression in brackets that contains the statistical factors for the four possible microscopic processes involved is the same as in our case, as we show next.

Let us start from our expression for the imaginary part of the (contracted) polarization tensor that leads to the rate, and take the limit $\Omega \rightarrow 0$ therein
\bea
\lim_{\Omega\rightarrow 0}\text{Im}\left[g_{\mu\nu}\Pi^{\mu\nu}_R(\omega)\right] &=& 4 q_f^2 \sum_{\sigma = \pm 1}\left( \lim_{\Omega\rightarrow 0}\text{Im}\left[\mathcal{I}_{\sigma}\right] + \lim_{\Omega\rightarrow 0}\text{Im}\left[\mathcal{J}_{\sigma}\right]\right)\nn\\
&=& -\pi q_f^2 m_f^2\int\frac{d^3k}{(2\pi)^3}\sum_{s=\pm 1} \Bigg\{\frac{\,n_\text{F}\left[\beta s E_k\right]}{ E_{kp} E_k}\left( \delta\left[ \wph - s\left(  E_k - E_{kp}\right) \right]-\delta\left[\wph -s\left( E_k + E_{kp}  \right) \right]\right)\nn\\
&-& \frac{\,n_\text{F}\left[\beta s E_{kp}\right]}{ E_k E_{kp}}\left( \delta\left[ \wph - s\left(  E_k - E_{kp}\right) \right] -
\delta\left[ \wph + s\left( E_{kp} + E_k \right)\right]\right)\Bigg\},
\label{eq_ratelim1}
\eea
where we used that $\lim_{\Omega\rightarrow 0}\text{Im}\mathcal{I}_{\sigma} = 0$ after Eq.~\eqref{eq_ImIsigma}, and we computed the corresponding limit $\lim_{\Omega\rightarrow 0}\text{Im}\mathcal{J}_{\sigma}$ from Eq.~\eqref{eq_ImJsigma}.
In order to compare with~\cite{Weldon_98_PhysRevD.28.2007}, let us switch to the notation $E_k = E_1$ and $E_{kp} = E_2$, such that after performing the sum over $s=\pm1$ explicitly in Eq.~\eqref{eq_ratelim1} we obtain
\bea
\lim_{\Omega\rightarrow 0}\text{Im}\left[g_{\mu\nu}\Pi^{\mu\nu}_R(\omega)\right] &=&  -\pi q_f^2 m_f^2\int\frac{d^3k}{(2\pi)^3} \Bigg\{\frac{\,n_\text{F}\left[\beta E_1\right]}{ E_1 E_2}\left( \delta\left[ \wph - E_1 + E_2 \right]-\delta\left[\wph -E_1 - E_2   \right]\right)\nn\\
&+& \frac{\,n_\text{F}\left[-\beta E_1\right]}{ E_1 E_2}\left( \delta\left[ \wph + E_1 - E_2 \right]-\delta\left[\wph +E_1 + E_2   \right]\right) - \frac{\,n_\text{F}\left[\beta E_2\right]}{ E_1 E_2}\left( \delta\left[ \wph - E_1 + E_2 \right]-\delta\left[\wph +E_1 + E_2   \right]\right)\nn\\
&-& \frac{\,n_\text{F}\left[-\beta E_2\right]}{ E_1 E_2}\left( \delta\left[ \wph + E_1 - E_2 \right]-\delta\left[\wph -E_1 - E_2   \right]\right) 
\Bigg\}.
\label{eq_ratelim1b}
\eea
Now, in order to compare with~\cite{Weldon_98_PhysRevD.28.2007}, let us change the notation to (for $j = 1,2$)
\bea
n_\text{F}\left[\beta E_j\right] &=& n_j\nn\\
n_\text{F}\left[-\beta E_j\right] &=& 1 - n_\text{F}\left[\beta E_j\right] = 1 - n_j,
\eea
and after factoring out the delta functions in Eq.~\eqref{eq_ratelim1b}, we arrive at
\bea
\lim_{\Omega\rightarrow 0}\text{Im}\left[g_{\mu\nu}\Pi^{\mu\nu}_R(\omega)\right] &=&  -\pi q_f^2 m_f^2\int\frac{d^3k}{(2\pi)^3} \frac{1}{E_1 E_2}\Bigg\{
\left( n_1 - n_2 \right)\delta(\omega - E_1 + E_2) - \left(1 - n_1 - n_2\right)\delta(\omega + E_1 + E_2)\nn\\
&+& \left(n_2 - n_1\right)\delta(\omega + E_1 - E_2) + \left(1 - n_2 - n_1\right)\delta(\omega - E_1 - E_2)\Bigg\}.
\label{eq_ratelim2}
\eea
Substituting the following elementary identities
\bea
n_1 - n_2 &=& n_1(1 - n_2) - n_2(1 - n_1)\nn\\
1 - n_1 - n_2 &=& (1 - n_1)(1 - n_2) - n_1 n_2
\eea
into Eq.~\eqref{eq_ratelim2}, we finally arrive at
\bea
\lim_{\Omega\rightarrow 0}\text{Im}\left[g_{\mu\nu}\Pi^{\mu\nu}_R(\omega)\right] &=&  -\pi q_f^2 m_f^2\int\frac{d^3k}{(2\pi)^3} \frac{1}{E_1 E_2}\Bigg\{
\delta(\omega - E_1 - E_2)\left[  (1 - n_1)(1 - n_2) - n_1 n_2\right]\nn\\
&-& \delta(\omega + E_1 - E_2)\left[n_1(1 - n_2) - n_2(1 - n_1) \right] - \delta(\omega-E_1 + E_2)\left[ n_2(1 - n_1) - n_1 (1 - n_2) \right]\nn\\
&+&\delta(\omega + E_1 + E_2)\left[ 
n_1 n_2 - (1 - n_1)(1 - n_2)\right]
\Bigg\},
\label{eq_ratelim3}
\eea
where it is evident by inspection that the integrand in curly brackets is the same as in the expression reported by Weldon~\cite{Weldon_98_PhysRevD.28.2007} as quoted in Eq.~\eqref{eq_Weldon}, as we claimed.
}

\end{document}